\documentclass[10pt,conference]{IEEEtran}
\IEEEoverridecommandlockouts
\usepackage[bottom,hang,flushmargin]{footmisc}
\usepackage[caption=false,font=footnotesize,textfont=normalfont]{subfig}

\renewcommand{\tablename}{Table}
\renewcommand{\figurename}{Figure}

\makeatletter
\long\def\@makecaption#1#2{%
  \vskip\abovecaptionskip
  \sbox\@tempboxa{{\normalsize\bfseries #1. #2}}%
  \ifdim \wd\@tempboxa >\hsize
    {\normalsize\bfseries #1. #2}\par
  \else
    \global \@minipagefalse
    \hb@xt@\hsize{\hfil\box\@tempboxa\hfil}%
  \fi
  \vskip\belowcaptionskip}
\makeatother

\usepackage{colortbl}
\usepackage{url}
\usepackage{xspace}
\usepackage{multirow}
\usepackage{adjustbox}
\usepackage[table,xcdraw]{xcolor}
\usepackage{tabularray}
\usepackage{array}
\usepackage{float}
\usepackage{amssymb}
\usepackage{pifont}
\usepackage{enumitem}
\usepackage{xstring}
\usepackage{bm}
\usepackage{makecell}
\usepackage{algorithm}
\usepackage[noend]{algorithmic}
\usepackage[normalem]{ulem}
\usepackage{microtype}
\usepackage{listings}
\usepackage{arydshln}

\definecolor{myred}{rgb}{0.6,0,0}
\definecolor{mygreen}{rgb}{0,0.6,0}
\definecolor{mymauve}{rgb}{0.58,0,0.82}
\definecolor{mediumgrayer}{gray}{0.85}
\usepackage[
  colorlinks=true,
  linkcolor=mygreen,
  citecolor=myred,
  urlcolor=blue
]{hyperref}

\usepackage[compress]{cite}
\usepackage{amsmath,amssymb,amsfonts}
\usepackage{graphicx}
\usepackage{textcomp}
\usepackage{booktabs}
\usepackage[most]{tcolorbox}

\def\BibTeX{{\rm B\kern-.05em{\sc i\kern-.025em b}\kern-.08em
    T\kern-.1667em\lower.7ex\hbox{E}\kern-.125emX}}

\newcommand*{\figu}{{Figure}\xspace}

\newcommand*{\tabl}{{Table}\xspace}

\newcounter{findingcounter}
\newcommand{\findingnum}{%
	\stepcounter{findingcounter}%
	\thefindingcounter}
    
\newcommand{\finding}{
    \textbf{Finding \findingnum:}
}

\makeatletter
\newcommand{\linebreakand}{%
  \end{@IEEEauthorhalign}
  \hfill\mbox{}\par
  \mbox{}\hfill\begin{@IEEEauthorhalign}
}
\makeatother

\begin{document}
%-------------------------------------------------------------------------------
\renewcommand{\footnoterule}{
  \vspace*{0.2cm} % 上方间距
  \hrule width 0.4\columnwidth
  \vspace*{0.2cm} % 下方间距
}

\definecolor{mygray}{rgb}{0.89, 0.89, 0.89}
\definecolor{lightgray}{gray}{0.8}

\title{Vulnerability-Affected Versions Identification: \\How Far Are We?
% {\footnotesize \textsuperscript{*}Note: Sub-titles are not captured for https://ieeexplore.ieee.org  and
}

\author{
\IEEEauthorblockN{
    Xingchu Chen\textsuperscript{1}, 
    Chengwei Liu\textsuperscript{2,3},
    Jialun Cao\textsuperscript{4},
    Yang Xiao\textsuperscript{1}\IEEEauthorrefmark{1}, \\
    Xinyue Cai\textsuperscript{1}, 
    Yeting Li\textsuperscript{1}, 
    Jingyi Shi\textsuperscript{1},
    Tianqi Sun\textsuperscript{1},
    Haiming Chen\textsuperscript{5},
    Wei Huo\textsuperscript{1}
}

\IEEEauthorblockA{
    \textsuperscript{1}Institute of Information Engineering, CAS; School of Cyber Security, UCAS, Beijing, China \\
    \textsuperscript{2}Nanyang Technological University, Singapore \\
    \textsuperscript{3}China-Singapore International Joint Research Institute (CSIJRI), Guangzhou, China \\
    \textsuperscript{4}The Hong Kong University of Science and Technology, Hong Kong, China \\
    \textsuperscript{5}Institute of Software, UCAS, Beijing, China \\
    \{chenxingchu, xiaoyang, caixinyue, liyeting, shijingyi, suntianqi, huowei\}@iie.ac.cn \\
    chengwei.liu@ntu.edu.sg, jcaoap@cse.ust.hk, chm@ios.ac.cn
}

    \thanks{\IEEEauthorrefmark{1}Corresponding author.}
}

\maketitle

%-------------------------------------------------------------------------------
\begin{abstract}
Identifying which software versions are affected by a vulnerability is critical for patching, risk mitigation. 
Despite a growing body of tools, their real-world effectiveness remains unclear due to narrow evaluation scopes—often limited to early SZZ variants, outdated techniques, and small or coarse-grained datasets.
In this paper, we present the first comprehensive empirical study of vulnerability-affected versions identification. We curate a high-quality benchmark of 1,128 real-world C/C++ vulnerabilities and systematically evaluate 12 representative tools from both tracing and matching paradigms across four dimensions: effectiveness at both vulnerability and version levels, root causes of false positives and negatives, sensitivity to patch characteristics, and ensemble potential.
Our findings reveal fundamental limitations: no tool exceeds 45.0\% accuracy, with key challenges stemming from heuristic dependence, limited semantic reasoning, and rigid matching logic. Patch structures such as add-only and cross-file changes further hinder performance. Although ensemble strategies can improve results by up to 10.1\%, overall accuracy remains below 60.0\%, highlighting the need for fundamentally new approaches.
Moreover, our study offers actionable insights to guide tool development, combination strategies, and future research in this critical area. Finally, we release the replicated code and benchmark on our website~\cite{vvstudy} to encourage future contributions.

\end{abstract}

%-------------------------------------------------------------------------------

%-------------------------------------------------------------------------------
\section{Introduction}

Software vulnerabilities continue to pose serious threats to system security. For defenders and researchers, it is critical to determine \textit{which software versions are affected by a known vulnerability}, referred to as \textit{vulnerability-affected version identification}. Accurate identification underpins a wide range of downstream applications, including vulnerability detection~\cite{Duan2017IdentifyingOL,Dong2024LibvDiffLV,Zhao2022OneBA,Zhao2022ALE}, automated patching~\cite{Yang2023EnhancingOP,Shariffdeen2021AutomatedPB}, and exploitability or propagation analysis~\cite{Liu2022DemystifyingTV,Zhang2023MitigatingPO,Wu2023UnderstandingTT}.

Despite its practical importance, reliably identifying affected versions remains challenging. Public vulnerability databases such as the NVD often contain incomplete or incorrect version metadata~\cite{Dong2019TowardsTD,Anwar2020CleaningTN,Wu2024VisionIA}, and many real-world vulnerabilities, especially those patched silently, are usually not documented at all~\cite{Zhou2023CoLeFunDaES,Zhou2021FindingAN,Luo2024StrengtheningSC}. As a result, practitioners frequently turn to analysis tools to infer affected versions retrospectively.

A range of such tools has emerged, falling broadly into two categories. Tracing-based methods~\cite{VccFinder,Lifetime,vszz,sem_szz,Vercation,tang2025llm4szz} extend the classic SZZ algorithm~\cite{Sliwerski2005WhenDC} to backtrack vulnerability-inducing commits and map them to released versions. Matching-based methods~\cite{sun2022verjava,Wu2024VisionIA,Shi2022PreciseV,woo2022movery,woo2023v1scan,mvp,kim2017vuddy} extract vulnerability-related code signatures and scan historical versions for semantic equivalence. However, despite promising individual results, the \textit{real-world effectiveness of these tools remains unclear} due to the lack of standardized, large-scale evaluation.
Previous studies fall short in several ways. First, existing evaluations focus narrowly on early tracing tools, assess only a few projects.~\cite{wen2019szzstudy,Rosa2021EvaluatingSI,lyu2024EvaluatingSI} And previous studies does not provide a thorough investigation. Second, inconsistent metrics and evaluation setups make cross-tool comparisons unreliable. Third, newer and more sophisticated tools, particularly in the matching domain, remain largely unevaluated. As a result, developers lack systematic guidance on tool selection or integration, and researchers cannot build on reliable baselines. To address this gap, our study goes substantially further by conducting stage-wise root cause analysis, quantifying the severity of failures, and identifying previously unreported challenges.

% Previous studies fall short in several ways. First, existing evaluations focus narrowly on early tracing tools, assess only a few projects, and yield coarse insights~\cite{wen2019szzstudy,Rosa2021EvaluatingSI,lyu2024EvaluatingSI}. Second, inconsistent metrics and evaluation setups make cross-tool comparisons unreliable. Third, newer and more sophisticated tools, particularly in the matching domain, remain largely unevaluated. As a result, developers lack systematic guidance on tool selection or integration, and researchers cannot build on reliable baselines.

Conducting a comprehensive evaluation presents two key challenges. First, building a high-quality benchmark requires curating ground-truth affected versions for real-world vulnerabilities, which is non-trivial given the inconsistent documentation, complex version histories, and the need for manual validation. Second, understanding why tools fail involves disentangling internal heuristics, code change semantics, and tool-specific assumptions, which are barely unveiled by prior studies.

\begin{figure}
    \centering
    \includegraphics[width=\linewidth]{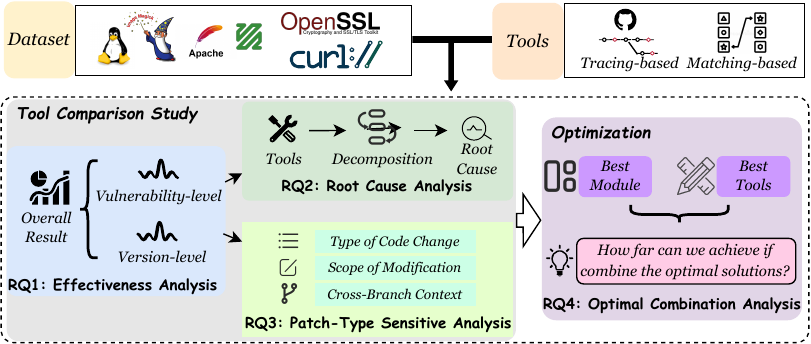}
    \setlength{\belowcaptionskip}{-10pt}
    \caption{Overview of Our Study}
    \label{fig:study_overview}
\end{figure}

To fill these gaps, we present the first systematic empirical study on vulnerability-affected versions identification, as shown in Figure~\ref{fig:study_overview}. 
First, we build a manually curated benchmark of 1,128 real-world vulnerabilities across 132 vulnerability types from diverse C/C++ projects. Created over 1.5 months with strict validation, this dataset provides fine-grained, reproducible ground truth for tool evaluation.
Second, we conduct a comprehensive comparison of 12 representative tools from both tracing and matching paradigms. The evaluation spans four dimensions: (1) effectiveness at both the vulnerability and version level, (2) root cause analysis of false positives (FPs) and false negatives (FNs), (3) sensitivity to diverse patch patterns (e.g., add-only, cross-file), and (4) potential gains from hybrid or ensemble strategies.
Third, to facilitate principled diagnosis, we introduce a stage-based decomposition framework that abstracts tracing and matching workflows into comparable steps. We then analyze the strategies of each tool at every stage, and conduct joint qualitative and quantitative assessments to pinpoint systematic errors.

Throughout the empirical study, we have discovered the following key findings: 1) no tool achieves more than 45.0\%  accuracy in identifying affected versions, raising serious concerns about their suitability for security-critical tasks; 2) The primary causes of FPs and FNs include heuristic over-reliance in tracing, insufficient semantic modeling and inflexible matching; 3) Add-only patches, cross-file changes, and multi-branch development significantly degrade performance; 4) Modular or ensemble combinations can improve accuracy by up to 10.1\%, but systemic design flaws cap accuracy below 60.0\%, signaling the need for fundamentally new approaches.

In summary, our main contributions are as follows:
\begin{itemize}[leftmargin=*]
    \item \textbf{Novelty} -- We constructed the first large-scale, high-quality benchmark for vulnerability-affected version identification, covering 1,128 vulnerabilities across 132 vulnerability types from diverse C/C++ projects. This manually curated dataset, built over 1.5 months, can effectively support further reproducible and reliable evaluation of identification tools. 
    \item \textbf{Rigorousness} -- We performed a comprehensive empirical evaluation of 12 representative tools from two major categories, offering an in-depth assessment of their effectiveness across multiple dimensions.
    \item \textbf{Significance} -- We identified key technical challenges and root causes behind tool limitations, and provided practical insights for tool improvement, ensemble design, and future research directions. 
    \item \textbf{Usefulness} -- We release the high-quality benchmark with 1,128 vulnerabilities across 132 vulnerability types from diverse C/C++ projects to ensure transparency and encourage future study. The artifact is released on the website~\cite{vvstudy}, including replicated code and the benchmark.

\end{itemize}

\section{Preliminaries and Related Work}

\subsection{Vulnerability-Affected Versions Identification}

Given a reported vulnerability $vul$ (e.g., a CVE) in a project $P$, public databases like NVD often list affected versions (e.g., via CPEs), but such metadata is frequently incomplete or inaccurate~\cite{Dong2019TowardsTD,Anwar2020CleaningTN,Wu2024VisionIA}, limiting its reliability for downstream tasks~\cite{Yang2023EnhancingOP,Zhao2022ALE}.
We define the task of \textbf{vulnerability-affected versions identification} as: given the set of released versions $V=\{v_1, v_2, ..., v_n\}$ of $P$, identify the subset $V_{\mathit{aff}} \subseteq V$ where each $v_i \in V_{\mathit{aff}}$ retains vulnerable logic of $vul$. Accurate identification of $V_{\mathit{aff}}$ is critical for managing N-day threats and enabling a timely response.

% Vulnerability-affected version detection aims to determine the range of software versions that are impacted by a known vulnerability. Formally, given a known vulnerability $v$ in a software project $P$, and a sequence of released versions $V = \{v_1, v_2, ..., v_n\}$, the task is to identify a subset $V_{\mathit{aff}} \subseteq V$ such that each $v_i \in V_{\mathit{aff}}$ contains the vulnerable code associated with $v$. 
% This task is critical for vulnerability management and remediation, as it helps assess which deployed versions require patching or mitigation.

% Generally, there are two major types of techniques for identifying affected versions of vulnerabilities in the literature, 1) inducing commit tracing based approaches built upon SZZ algorithm~\cite{Sliwerski2005WhenDC}, and 2) directly adopting vulnerability detection approaches to determine the existence of vulnerabilities in versions. 

% \subsubsection{\textbf{Inducing Commit Tracing based Approaches}}

% Some researchers borrows the SZZ algorithm~\cite{Sliwerski2005WhenDC} from bug detection to identify the inducing commit of vulnerabilities so that determining the affected versions. As far as we know, only V-SZZ\cite{vszz} used the SZZ algorithm to identify the inducing commit of vulnerability. They specializes in mapping deleted lines by iterative tracing backward until the codes are no longer similar. And choose the oldest candidate commit as inducing commit.
% \cxc{Only V-SZZ identify vulnerability inducing commit. LLM4SZZ only identify bug-inducing commit in papers}

\subsection{Existing Identification Approaches}

Generally, existing approaches for identifying vulnerability-affected versions fall into two main types. Tracing-based methods build on the SZZ algorithm~\cite{Sliwerski2005WhenDC} to locate vulnerability-inducing commits and infer affected versions. Matching-based methods directly identify vulnerable logic across versions through static techniques (e.g., syntactic or semantic pattern matching) or dynamic analysis (e.g., fuzzing).

\subsubsection{\textbf{Tracing-based Approaches}}
Tracing-based methods infer affected versions by first identifying the commits that introduced a vulnerability. A number of approaches build on the SZZ algorithm~\cite{Sliwerski2005WhenDC}, originally designed for bug-inducing commit identification.
\texttt{VCCFinder}~\cite{VccFinder} was the first to adapt SZZ to vulnerability scenarios, while \texttt{Lifetime}~\cite{Lifetime} improved commit identification with fine-grained heuristics. \texttt{V-SZZ}~\cite{vszz} introduced the idea of mapping vulnerability-inducing commits to version tags, enabling affected version inference.

Numerous SZZ variants (e.g., \texttt{AG-SZZ}~\cite{Kim2006AutomaticIO}, \texttt{MA-SZZ}~\cite{maszz}, \texttt{RA-SZZ}~\cite{Neto2018Raszzstudy}, \texttt{PR-SZZ}~\cite{Bludau2022PRszz}) have been proposed to improve the precision of bug-inducing commit identification. 
More recent efforts explore semantic and learning-based enhancements. \texttt{Neural-SZZ}~\cite{Neural_SZZ} identifies vulnerable code using graph neural networks. \texttt{Vercation}~\cite{Vercation} and \texttt{Sem-SZZ}~\cite{sem_szz} apply program slicing for semantic backtracking. \texttt{LLM4SZZ}~\cite{tang2025llm4szz} leverages large language models (LLMs) to refine both root cause localization and commit selection.

\subsubsection{\textbf{Matching-based Approaches}}
% \noindent 2) \textbf{\textit{Matching-based Approaches}}: 

The matching-based tools are mainly designed for recurring vulnerability detection. While tools like \texttt{ReDeBug}~\cite{jang2012redebug} were originally developed for clone detection, others such as \texttt{VISION}~\cite{Wu2024VisionIA} and \texttt{VerJava}~\cite{sun2022verjava} extend this idea specifically to identify affected versions which aligns with our evaluated scenario. These approaches evolve along multiple axes—ranging from syntactic matching to semantic analysis and structural representations.
Early techniques, such as \texttt{ReDeBug}~\cite{jang2012redebug}, Li et al.~\cite{li2014scalable}, \texttt{PatchGen}\cite{luo2015PatchGen} and \texttt{VUDDY}~\cite{kim2017vuddy}, relied on syntactic signatures or token-level features, offering fast yet coarse-grained matching.

To enhance accuracy and reduce false positives, later methods incorporated semantic context using program slicing, taint analysis, and version filtering. Representative tools include \texttt{MVP}~\cite{mvp}, \texttt{MOVERY}~\cite{woo2022movery}, \texttt{Tracer}~\cite{woo2022Tracer}, \texttt{V1SCAN}~\cite{woo2023v1scan}, \texttt{FIRE}~\cite{feng2024fire}, and \texttt{VULTURE}~\cite{vulture}.
To further generalize matching across codebases and abstractions, structural approaches emerged. These include AST-based models like \texttt{VMUD}~\cite{Huang2024Vmud}, \texttt{PATEN}~\cite{Li2025Paten}, and \texttt{VISION}~\cite{Wu2024VisionIA}, as well as graph-based representations such as \texttt{HiddenCPG}~\cite{Wi2022HiddenCPG}. These techniques enable richer modeling of vulnerability patterns and greater resilience to code modifications.

Besides, dynamic methods~\cite{Dai2021FacilitatingVA,Zhang2024SymBisectAB} use symbolic execution or proof-of-concept (PoC) exploits to validate vulnerability presence in each version. While highly precise, such methods face practical limitations due to the unavailability of PoCs and challenges in reliably building and executing historical versions. So they 
% their applicability 
remain constrained in large-scale, automated settings.

\subsection{Existing Evaluations}
Several studies have examined the effectiveness of SZZ and its variants to understand their limitations and improve their design. 
Wen et al.~\cite{wen2019szzstudy} highlighted that SZZ often fails when there is no direct overlap between the fixing and inducing commits. Rezk et al.~\cite{Rezk2022GhostStudy} investigated the impact of ghost commits on the SZZ algorithm and construct the evaluation dataset with developers' information. Rosa~\cite{Rosa2021EvaluatingSI} introduced an NLP-based method for automatically constructing evaluation datasets, while Lyu et al.~\cite{lyu2024EvaluatingSI} investigated the influence of ghost commits and explored the theoretical upper bound of SZZ's accuracy. 
However, these evaluations primarily focus on early SZZ variants, use limited or unverified datasets, and cover only a small number of projects. As a result, they offer only a partial view of the problem and fall short of revealing broader tool effectiveness, robustness, or real-world applicability—gaps that our comprehensive empirical study aims to fill.

% Some researchers have conducted comparative studies to investigate the pros and cons of these SZZ derivatives, to better optimize the algorithm. 
% Wen et al.~\cite{wen2019szzstudy} pointing out that SZZ algorithms struggle when there's no direct overlap between the fixing commit and the inducing commit. Rosa~\cite{Rosa2021EvaluatingSI} proposed a method for automatically constructing SZZ evaluating datasets using Natural Language Processing (NLP). Lyu et al.~\cite{lyu2024EvaluatingSI} primarily analyzed the impact of ghost commits on the SZZ algorithm and provided valuable insights into exploring its upper limits. These studies focused narrowly on early SZZ variants, evaluated a few projects with automatically crafted dataset, and reached simplified conclusions.

\subsection{Motivation of this paper}

Considering the exponential growth in exploited vulnerabilities, the accurate management of vulnerability information, particularly the precise delineation of affected versions, has become more critical than ever for downstream tasks, such as patch deployment and risk evaluation. Unfortunately, current vulnerability datasets often exhibit low quality and inaccurate metadata, leaving practitioners without the detail required for effective remediation. Although existing works have been proposed in the fields of inducing commit tracing based approaches and vulnerability matching based detections, none has systematically investigated how well these two types of solutions support the automated identification of vulnerable versions in real‑world settings. To address this gap, in this paper, we conduct the first comprehensive study that evaluates their practical capabilities and clarifies how far we have progressed toward fully automated, accurate vulnerable version identification, thereby illuminating directions for future research.

\section{Study Design}\label{overview}

% \input{table/overview_tools_details}

% In this section, we present a comprehensive and accurately curated dataset along with its construction methodology. This dataset is not only applicable for vulnerable version evaluation but also serves as a valuable resource for recurring vulnerability detection and related research areas.

\begin{table}[!tb]
\caption{List of selected tools. \# Baseline: Tool used as a baseline. \# Citation: Number of citations.}
\label{tab:tool_overview}
    \centering
    \renewcommand\arraystretch{1.1}
    \setlength{\tabcolsep}{1.3mm}
    \resizebox{0.95\linewidth}{!}{
    \begin{tabular}{clrrr}
    \toprule
        \textbf{Type} & \textbf{Tool} & \textbf{\#Baseline} & \textbf{\#Citation} & \textbf{Publication} \\ 
        % \hline
        \midrule

        \multirow{6}[1]{*}{\textbf{Tracing-based}} & 
        \cellcolor{mygray}VCCFinder & \cellcolor{mygray}7 & \cellcolor{mygray}330 & \cellcolor{mygray}CCS'15 \\

        & V-SZZ & 6 & 50 & ICSE'22 \\

        & \cellcolor{mygray}Lifetime & \cellcolor{mygray}0 & \cellcolor{mygray}33 & \cellcolor{mygray}SEC'22 \\

        & SEM-SZZ & 0 & 0 &  TSE'24 \\

        & \cellcolor{mygray}TC-SZZ & \cellcolor{mygray}0 & \cellcolor{mygray}3 & \cellcolor{mygray}TSE'24 \\

        % & VERCATION & 0 & 7 & arXiv'24~\cite{Vercation} \\

        & LLM4SZZ & 0 & 0 & ISSTA'25\\ 
        
        % \hline
        \midrule

         \multirow{6}[1]{*}{\textbf{Matching-based}} &

         \cellcolor{mygray}ReDeBug & \cellcolor{mygray}24 & \cellcolor{mygray}314 & \cellcolor{mygray}S\&P'12 \\

         & VUDDY & 47 & 467 & S\&P'17 \\

         % & MVP & 7 & 126 & SEC'20~\cite{mvp} \\

         & \cellcolor{mygray}MOVERY & \cellcolor{mygray}3 & \cellcolor{mygray}39 & \cellcolor{mygray}SEC'22 \\

         & V1SCAN & 3 & 16 & SEC'23 \\

         & \cellcolor{mygray}FIRE & \cellcolor{mygray}0 & \cellcolor{mygray}2 & \cellcolor{mygray}SEC'24 \\

         & VULTURE & 0 & 0 & NDSS'25 \\ 
         \bottomrule
    \end{tabular}
    }
\end{table}

\subsection{Tool Selection}
\label{tool_selection}
We first conduct a systematic literature review (SLR) grounded in rigorous criteria to collect related papers on these two types of approaches.
% To select representative tools for evaluating tracing- and matching-based approaches to identifying vulnerability-affected versions, we conducted a systematic literature review (SLR) grounded in rigorous criteria.
Specifically, we target at top-tier software engineering and security venues (e.g., ICSE, ASE, IEEE S\&P, USENIX Security) to search related papers published between May 2020 and May 2025, using keywords such as “vulnerable version”, “version range”, “SZZ”, and “recurring vulnerability.” This yielded 22 papers. After that, we further apply backward and forward snowballing to identify other relevant papers that are missed by keyword searching, and another 19 papers are included, resulting in 41 relevant papers in total for tool selection.

As most of these techniques target C/C++ projects, which is the most concerned ecosystem in vulnerability research, we excluded tools specific to other environments (e.g., \texttt{VerJava}~\cite{sun2022verjava}, \texttt{Neural-SZZ}~\cite{Neural_SZZ}, \texttt{AFV}~\cite{Shi2022PreciseV}) to maximize comparison targets.
After that, we further filter out papers that are (1) not proposing new tools, (2) tools not being available(i.e.,\texttt{VISION}~\cite{Wu2024VisionIA}), and (3) requiring additional information (i.e., \texttt{OpenSZZ}~\cite{pellegrini2019openszz} requires additional information from Jira issues~\cite{Jira}) or compilation (i.e., \texttt{Tracer}~\cite{woo2022Tracer}) that are infeasible for automated identification of vulnerable versions for general vulnerabilities. 
% We further refined the selection to tools that are (1) publicly available and open-source, and (2) representative, either by academic influence (e.g., 10+ citations for pre-2023 works), or by recent recognition (e.g., publication at a top-tier venue in 2024–2025), even if citation and GitHub metrics are still emerging. This hybrid criterion balances established impact with timely inclusion of recent advances.
As a result, 12 tools in total, 6 matching-based tools and 6 tracing-based tools, covering diverse methodologies including heuristics, semantic reasoning, and LLM-powered analysis, are finally selected as the target tools of this study. Table~\ref{tab:tool_overview} summarizes the selected tools, their classification, citation status, and publication venues. The detailed table of the SLR is presented on our website~\cite{vvstudy}. 

\begin{table*}[!htb]
    \centering
    \caption{\textbf{Overview of the Constructed Dataset.} 
\#Star: GitHub stars; \#CVE: collected CVEs; \#Patch: total patches; 
\#Add-only / \#Del-only / \#Mixed: patch types by diff composition; 
\#Version: total affected versions; Branch: development model.}
    \label{tab:dataset_details}
    \setlength{\tabcolsep}{1.3mm}
    \resizebox{0.9\textwidth}{!}{
    \begin{tabular}{lcllllclll}
    \toprule
        \textbf{Project} & \textbf{\#Star} & \textbf{Domain} & \textbf{\#CVE} & \textbf{\#Patch} & \textbf{\#Add-only}& \textbf{\#Del-only}& \textbf{\#Mixed} & \textbf{\#Version} & \textbf{Branch}\\ \midrule
        \rowcolor[rgb]{0.89, 0.89, 0.89} Linux kernel & 190k & Operating Systems & 717 & 928 & 222 & 17 & 689 & 17,324 & Multiple\\        
        FFmpeg & 49k & Multimedia Processing & 71 & 238 & 35 & 0 & 203 & 9,428 & Multiple\\
        \rowcolor[rgb]{0.89, 0.89, 0.89} Curl & 37k & Command-line Data Transfer & 68 & 66 & 12 & 1 & 53 & 4,601 & Single\\
        OpenSSL & 27k & Cryptography and Secure Communication & 50 & 50 & 16 & 1 & 33 & 1,796 & Multiple\\
        \rowcolor[rgb]{0.89, 0.89, 0.89}ImageMagick & 13k & Image Processing & 72 & 71 & 7 & 0 & 64 & 12,377 & Single\\  
        QEMU & 11k & System Virtualization and Emulation & 57 & 78 & 27 & 1 & 50 & 2,188 & Multiple\\
        \rowcolor[rgb]{0.89, 0.89, 0.89}Wireshark & 8k & Network Traffic Analysis & 50 & 53 & 4 & 0 & 49 & 8,849 & Multiple\\
        Apache HTTP Server & 4k & Web Server & 30 & 43 & 2 & 0 & 41 & 2,496 & Single\\
       \rowcolor[rgb]{0.89, 0.89, 0.89} OpenJPEG & 1k & Image Processing & 13 & 15 & 4 & 0 & 11 & 128 & Single \\
        \midrule
        Total & - & - & 1,128 & 1,542 & 329 & 20 & 1,193 & 59,187 & - \\
        
    \bottomrule
    \end{tabular}
    }
\end{table*}

\subsection{Dataset Construction}
\label{dataset_construction}

Existing datasets used for evaluating vulnerable version identification tools suffer from several key limitations: they either (1) target a single project~\cite{he2023autoidentifyVersion,lyu2024EvaluatingSI} (e.g., the Linux kernel), which include a limited number of vulnerabilities (e.g., only 100 in \texttt{V-SZZ}~\cite{vszz}), or (2) are with poor accuracy in their ground truths (\texttt{Lifetime}~\cite{Lifetime} directly takes the labeled version in NVD, which is known as inaccurate).
% contain labeling errors (e.g., incorrect affected version ranges in the Vercation dataset for most FFmpeg vulnerabilities~\cite{Vercation}). \todo{any other limitations?} 
These issues hinder the generalizability and reliability of evaluation results.
To address these limitations, we construct a new benchmark that is large-scale, diverse, and accurately annotated. The construction process consists of three main steps:

\textit{\textbf{1) Project Selection.}} 
We select nine representative C/C++ projects based on the following criteria: (i) \textit{popularity}, measured by GitHub stars, forks, and prior usage in vulnerability research; (ii) \textit{vulnerability density}, ensuring the selected projects have a substantial number of reported CVEs; and (iii) \textit{domain diversity}, covering a range of application areas.

\textit{\textbf{2) Patch Collection.}}
For each selected project, we manually collect fixing commit of vulnerability from Jan 2020 to Dec 2024.
% \bella{Could we add a specific date range here?}. 
Patch identification is primarily based on CVE references from NVD,
% \bella{Should we add a link here?}
and is cross-referenced with official security advisories or project-maintained vulnerability logs.

\textit{\textbf{3) Ground Truth Annotation.}}
We determine affected version ranges following the annotation methodology of \texttt{V-SZZ}~\cite{vszz}. Then we further validated the affected versions as follows:

% \bella{What is the `enhanced validation'? Do we mean the following statements? Then can we say `...VSZZ. Then, we further validated the detected versions as follows:'?}. 

% \bella{Recently, I found that reviewers like to ask about the authority of the annotators. So, can we say something like `... annotators with xx-year experience in ...'?} 
Two annotators independently identify vulnerable statements by consulting multiple sources, including CVE descriptions, vulnerability reports, GitHub issues and fixing commit. They trace the commit history of the vulnerable statements to identify the vulnerability-inducing commit. All versions between the inducing-commit and patch-commit are labeled as vulnerable. Disagreements are resolved by a third annotator through discussion and evidence inspection. Each annotator has 8 years experience in C/C++ language programming and vulnerability analysis. This multi-stage process ensures high labeling accuracy and mitigates individual bias. Among all vulnerabilities, 134(11.9\%) vulnerabilities had initial annotation inconsistencies, and the Cohen’s Kappa for the labeling is 0.83.

On average, annotating each CVE requires approximately \textbf{0.5 person-hours}. To reduce complexity and preserve evaluation fidelity, we excluded release-candidate (RC) builds while labeling versions. The dataset can be found in our artifact. 
Table~\ref{tab:dataset_details} summarizes key statistics of our constructed dataset, which includes affected version annotations for \textbf{1,128 CVEs}, covering a total of \textbf{59,983 vulnerable versions}. The dataset spans \textbf{132 distinct CWE types}, with broad coverage across vulnerability categories. To the best of our knowledge, this is the largest publicly available benchmark for vulnerability-affected version identification.
% \bella{Can we add these statistics in the second contribution?}

\subsection{Experimental Setup}

We evaluate tools under default settings. Tracing-based tools follow \texttt{V-SZZ}~\cite{vszz}, labeling versions between introducing and fixing commits as vulnerable. For \texttt{LLM4SZZ}, we use {llama3.1-70b}~\cite{tang2025llm4szz}. 
For matching-based tools, we provide the fixing commit as input and apply each tool to the entire codebase of all historical versions. These tools extract features based on the patch and search for similar features to determine whether the version is vulnerable. We strictly follow each tool’s original setup to ensure correctness.

\section{Comparison and Evaluation}\label{evaluation}

% In this section, we present the result of experiment, accompanied by corresponding \textbf{Finding}. 
% Here, observation refer to factual and objective findings, while insights represent deeper understandings that can guide researchers and developers in applying and improving vulnerable version detection tools.

The evaluation answers four research questions (RQs):

\begin{itemize}
    \item \textbf{RQ1: Effectiveness Analysis.} \textit{How effective are current tools in identifying affected versions of vulnerabilities?}
    \item \textbf{RQ2: Root Cause Analysis.} \textit{What are the primary causes of FPs and FNs produced by existing tools?}
    \item \textbf{RQ3: Patch-Type Sensitivity Analysis.} \textit{How does identification performance vary across different patch types?}
    \item \textbf{RQ4: Tool Combination Analysis.} \textit{Can combining existing tools improve the overall effectiveness?}
\end{itemize}

\subsection{Effectiveness Analysis (RQ1)}
\label{rq1}

\begin{table*}[htbp]
\centering
\caption{Effectiveness of tools at the vulnerability and version levels. NM: No-Miss; NMR: No-Miss Ratio.}
\label{tab:overall_results}%
\renewcommand{\arraystretch}{1.3}
\resizebox{0.9\textwidth}{!}{
\begin{tabular}{lccccccccccc}

    \toprule
    \multicolumn{1}{c}{\multirow{2}[2]{*}{\textbf{Type}}} & \multicolumn{1}{c}{\multirow{2}[2]{*}{\textbf{Tool}}} & \multicolumn{4}{c}{\textbf{Vulnerability-level}}             & \multicolumn{6}{c}{\textbf{Version-level}} \\
    \cmidrule(lr){3-6} \cmidrule(lr){7-12}    
    \multicolumn{1}{c}{} &       & \textbf{TP} & \textbf{Accuracy} & \textbf{NM} & \textbf{NMR} & \textbf{FP} & \textbf{FN} & \textbf{TP} & \textbf{Precision} & \textbf{Recall} & \textbf{F1} \\
    \cmidrule(lr){1-6} \cmidrule(lr){7-12} 
    \multirow{6}[1]{*}{\textbf{Tracing-based}} 
    & \cellcolor[rgb]{0.89, 0.89, 0.89}\textbf{VCCFinder} & \cellcolor[rgb]{0.89, 0.89, 0.89}\textbf{506} & \cellcolor[rgb]{0.89, 0.89, 0.89}\textbf{44.9\%} & \cellcolor[rgb]{0.89, 0.89, 0.89}\textbf{776} & \cellcolor[rgb]{0.89, 0.89, 0.89}\textbf{68.8\%} & \cellcolor[rgb]{0.89, 0.89, 0.89}12020 & \cellcolor[rgb]{0.89, 0.89, 0.89}13879 & \cellcolor[rgb]{0.89, 0.89, 0.89}45308 & \cellcolor[rgb]{0.89, 0.89, 0.89}79.0\% & \cellcolor[rgb]{0.89, 0.89, 0.89}76.6\% & \cellcolor[rgb]{0.89, 0.89, 0.89}\textbf{77.8\%} \\
    &  \textbf{V-SZZ} &  468   & 41.5\% & 768   & 68.1\% & 16086 & 12215 & 46972 & 74.5\% & \textbf{79.4\%} & 76.8\% \\
    & \cellcolor[rgb]{0.89, 0.89, 0.89}\textbf{Lifetime}  & \cellcolor[rgb]{0.89, 0.89, 0.89}505   & \cellcolor[rgb]{0.89, 0.89, 0.89}44.8\% & \cellcolor[rgb]{0.89, 0.89, 0.89}773   & \cellcolor[rgb]{0.89, 0.89, 0.89}68.5\% & \cellcolor[rgb]{0.89, 0.89, 0.89}11974 & \cellcolor[rgb]{0.89, 0.89, 0.89}13922 & \cellcolor[rgb]{0.89, 0.89, 0.89}45265 & \cellcolor[rgb]{0.89, 0.89, 0.89}79.1\% & \cellcolor[rgb]{0.89, 0.89, 0.89}76.5\% & \cellcolor[rgb]{0.89, 0.89, 0.89}\textbf{77.8\%} \\
    &  \textbf{SEM-SZZ} & 463   & 41.0\% & 621   & 55.1\% & 6257  & 21375 & 37812 & 85.8\% & 63.9\% & 73.2\% \\
    & \cellcolor[rgb]{0.89, 0.89, 0.89}\textbf{TC-SZZ} & \cellcolor[rgb]{0.89, 0.89, 0.89}195   & \cellcolor[rgb]{0.89, 0.89, 0.89}17.3\% & \cellcolor[rgb]{0.89, 0.89, 0.89}518   & \cellcolor[rgb]{0.89, 0.89, 0.89}45.9\% & \cellcolor[rgb]{0.89, 0.89, 0.89}24356 & \cellcolor[rgb]{0.89, 0.89, 0.89}23719 & \cellcolor[rgb]{0.89, 0.89, 0.89}35468 & \cellcolor[rgb]{0.89, 0.89, 0.89}59.3\% & \cellcolor[rgb]{0.89, 0.89, 0.89}59.9\% & \cellcolor[rgb]{0.89, 0.89, 0.89}59.6\% \\
    &  \textbf{LLM4SZZ}  & 459   & 40.7\% & 664   & 58.9\% & 9491  & 20281 & 38906 & 80.4\% & 65.7\% & 72.3\% \\
    &  \cellcolor[rgb]{0.89, 0.89, 0.89} \textit{\textbf{Sum}} & \cellcolor[rgb]{0.89, 0.89, 0.89}\textit{\textbf{2,596}} & \cellcolor[rgb]{0.89, 0.89, 0.89}\textit{\textbf{38.4\%}} & \cellcolor[rgb]{0.89, 0.89, 0.89}\textit{\textbf{4120}} & \cellcolor[rgb]{0.89, 0.89, 0.89}\textit{\textbf{60.9\%}} & \cellcolor[rgb]{0.89, 0.89, 0.89}\textit{\textbf{80,184}} & \cellcolor[rgb]{0.89, 0.89, 0.89}\textit{\textbf{105,391}} & \cellcolor[rgb]{0.89, 0.89, 0.89}\textit{\textbf{249,731}} & \cellcolor[rgb]{0.89, 0.89, 0.89}\textit{\textbf{75.7\%}} & \cellcolor[rgb]{0.89, 0.89, 0.89}\textit{\textbf{70.3\%}} & \cellcolor[rgb]{0.89, 0.89, 0.89}\textit{\textbf{72.9\%}} \\
    \cmidrule(lr){1-6} \cmidrule(lr){7-12}
    \multirow{6}[1]{*}{\textbf{Matching-based}} & 
     \textbf{ReDeBug} & 417   & 37.0\% & 536   & 47.5\% & 3989  & 23289 & 35898 & 90.0\% & 60.7\% & 72.5\% \\
    & \cellcolor[rgb]{0.89, 0.89, 0.89} \textbf{VUDDY} & \cellcolor[rgb]{0.89, 0.89, 0.89} 243 & \cellcolor[rgb]{0.89, 0.89, 0.89} 21.5\% & \cellcolor[rgb]{0.89, 0.89, 0.89} 296 & \cellcolor[rgb]{0.89, 0.89, 0.89} 26.2\% & \cellcolor[rgb]{0.89, 0.89, 0.89} 1227 & \cellcolor[rgb]{0.89, 0.89, 0.89} 39426 & \cellcolor[rgb]{0.89, 0.89, 0.89} 19761 & \cellcolor[rgb]{0.89, 0.89, 0.89} \textbf{94.2\%} & \cellcolor[rgb]{0.89, 0.89, 0.89} 33.4\% & \cellcolor[rgb]{0.89, 0.89, 0.89} 49.3\% \\

    & \textbf{MOVERY} & 374   & 33.2\% & 622   & 55.1\% & 11604 & 18642 & 40545 & 77.7\% & 68.5\% & 72.8\% \\
    & \cellcolor[rgb]{0.89, 0.89, 0.89} \textbf{V1SCAN} & \cellcolor[rgb]{0.89, 0.89, 0.89} 326 & \cellcolor[rgb]{0.89, 0.89, 0.89} 28.9\% & \cellcolor[rgb]{0.89, 0.89, 0.89} 424 & \cellcolor[rgb]{0.89, 0.89, 0.89} 37.6\% & \cellcolor[rgb]{0.89, 0.89, 0.89} 2692 & \cellcolor[rgb]{0.89, 0.89, 0.89} 26719 & \cellcolor[rgb]{0.89, 0.89, 0.89} 32468 & \cellcolor[rgb]{0.89, 0.89, 0.89} 92.3\% & \cellcolor[rgb]{0.89, 0.89, 0.89} 54.9\% & \cellcolor[rgb]{0.89, 0.89, 0.89} 68.8\% \\

    &  \textbf{FIRE} & 406   & 36.0\% & 517   & 45.8\% & 4316  & 23236 & 35951 & 89.3\% & 60.7\% & 72.3\% \\
    & \cellcolor[rgb]{0.89, 0.89, 0.89} \textbf{VULTURE} & \cellcolor[rgb]{0.89, 0.89, 0.89} 44 & \cellcolor[rgb]{0.89, 0.89, 0.89} 3.9\% & \cellcolor[rgb]{0.89, 0.89, 0.89} 622 & \cellcolor[rgb]{0.89, 0.89, 0.89} 55.1\% & \cellcolor[rgb]{0.89, 0.89, 0.89} 54889 & \cellcolor[rgb]{0.89, 0.89, 0.89} 28063 & \cellcolor[rgb]{0.89, 0.89, 0.89} 31124 & \cellcolor[rgb]{0.89, 0.89, 0.89} 36.2\% & \cellcolor[rgb]{0.89, 0.89, 0.89} 52.6\% & \cellcolor[rgb]{0.89, 0.89, 0.89} 42.9\% \\
    &  \textit{\textbf{Sum}} & \textit{\textbf{1,810}} & \textit{\textbf{26.7\%}} & \textit{\textbf{3,017}} & \textit{\textbf{44.6\%}} & \textit{\textbf{78,717}} & \textit{\textbf{159,375}} & \textit{\textbf{195,747}} & \textit{\textbf{71.3\%}} & \textit{\textbf{55.1\%}} & \textit{\textbf{62.2\%}} \\ 
    \bottomrule 
    
    \end{tabular}
}
\end{table*}

% \input{table/trace_match_compare_table}

% \subsubsection{Setup}

% We evaluate the effectiveness of tracing-based and matching-based tools from two complementary perspectives: \textit{vulnerability level} and \textit{version level}.

% At the \textbf{vulnerability level}, the goal is to assess whether a tool can correctly identify all affected versions for each vulnerability. A prediction is counted as \textit{correct} (i.e., true positive) only if it exactly matches the ground truth—containing no false positives or false negatives. Additionally, we define a \textit{no-miss} case as one where all true affected versions are included, even if extra versions are reported. This distinction reflects practical trade-offs: missing affected versions may compromise security, while over-reporting mainly incurs additional maintenance effort.

% At the \textbf{version level}, each version prediction is evaluated independently to capture partial correctness and better characterize a tool's generalization. Tools may produce both false positives and false negatives, revealing different bias patterns.

% We use two metrics at the vulnerability level: \textit{accuracy}, defined as $\text{Accuracy} = \frac{\#\text{TP}}{\#\text{GT}}$, and \textit{no-miss ratio}, defined as $\text{No-Miss Ratio} = \frac{\#\text{No-Miss}}{\#\text{GT}}$, where $\#\text{GT}$ denotes the total number of vulnerabilities. At the version level, we report standard metrics—\textit{precision}, \textit{recall}, and \textit{F1-score}—to quantify performance over individual version predictions.

\subsubsection{\textbf{Setup}}

The effectiveness
% of tracing-based and matching-based methods 
is evaluated from two complementary perspectives:

\begin{itemize}[leftmargin=*]
    \item \textbf{Vulnerability-level Evaluation.} This evaluation assesses whether a tool accurately identifies the full set of affected versions for each vulnerability. A prediction is considered \textit{correct} (true positive) only if it exactly matches the ground truth (no missing or extra versions). We also define a \textit{no-miss} case as one that contains all ground-truth versions, despite the extra ones. This distinction reflects practical trade-offs: missing affected versions may introduce security risks, while over-reporting primarily increases maintenance overhead.
    We report two metrics: \textit{Accuracy} ($\text{TP} / \text{Total}$), and \textit{No-Miss Ratio} (NMR = $\text{No-Miss} / \text{Total}$), where \textit{Total} is the number of evaluated vulnerabilities.

  \item \textbf{Version-level Evaluation.} Each version is evaluated independently to capture partial correctness and better reflect a tool’s generalization ability. Tools may yield both FNs and FNs, reflecting different error tendencies. We adopt standard metrics at this level: \textit{precision}, \textit{recall}, and \textit{F1-score}.
\end{itemize}

\subsubsection{\textbf{Results}}

\tabl~\ref{tab:overall_results} summarizes the effectiveness of each tool at both the vulnerability and version levels.

\textbf{Overall Results.}
Among all tools, \texttt{VCCFinder} achieves the highest performance, with an accurace of 44.9\%, a No-Miss Ratio of 68.8\%, and an F1-score of 77.8\%. Only five tools achieve an accuracy of at least 40.0\%, and only three tools reach a No-Miss Ratio above 60.0\%. These results highlight a significant deficiency in the current methods' ability to accurately identify the complete set of affected versions. 
% Even the best tool (i.e., \texttt{VCCFinder}) fails to provide reliable completeness for practical use \bella{Could we say `Even the accuracy of the best tool is less than half, which is far beyond a reliable usage in the practice'?}.
Even the accuracy of the best tool (i.e., \texttt{VCCFinder}) is less than half, which is far beyond a reliable usage in the practice.
Moreover, all evaluated tools miss part of the affected versions for at least 30.0\% of the vulnerabilities, which posing substantial security risks to downstream software leading to potentially unpatched vulnerabilities. 
% While all 12 tools show higher F1-scores at the version level, the incomplete affected version identification still hinder the use of existing methods.

% such a partial correctness does not ensure full vulnerability coverage.

% \textbf{Tracing-based Methods.}
% Tracing-based tools exhibit diverse performance. At the vulnerability level, VCCFinder achieves the highest F1-score (51.6\%), while SEM-SZZ attains the best precision (68.9\%) but suffers from lower recall (41.0\%). In contrast, TC-SZZ perform poorly, with F1-scores below 25\%, largely due to high false positive rates and low recall.
% Performance improves at the version level. VCCFinder and Lifetime achieve the best F1-score (77.8\%), while SEM-SZZ delivers excellent precision (85.8\%) at the cost of reduced recall (63.9\%). 

% \textbf{Matching-based Methods.}
% Among matching-based tools, ReDeBug achieves the highest F1-score at the vulnerability level (49.3\%), with high precision (74.1\%) and moderate recall (37.0\%). VUDDYexhibits even higher precision (78.9\%) but suffers from low recall (21.5\%), reflecting its conservative matching strategy. FIRE and V1SCAN offer more balanced performance, both achieving F1-scores above 40\%.
% At the version level, all tools show notable improvements. MOVERY leads with the highest F1-score (72.8\%) and strong recall (68.5\%), followed by ReDeBug and FIRE. Despite its early design, ReDeBug continues to deliver competitive performance.

\begin{tcolorbox}[colback=black!10!white,enhanced,frame hidden, boxsep=0pt,left=5pt,right=5pt,top=2pt,bottom=2pt]
\finding
Existing tools remain limited in accurately inferring affected versions. The best tool achieves under 50.0\% accuracy at the vulnerability level, with over 30.0\% of vulnerabilities missing at least one affected version.
% Existing tools show significant limitations in accurately identifying the complete set of affected versions, with even the best-performing tool achieving less than 50\% accuracy at the vulnerability level. In over 30\% of the cases, the tools fail to identify all affected versions of a vulnerability.
% \bella{Can we also highlight NMR here? For example, we can say `Less than 70\% of the affected versions of the vulnerability can be fully identified'.}
\end{tcolorbox}

\textbf{Vulnerability-Level vs. Version-Level Discrepancy.}
We observe a consistent and significant gap between evaluation metrics at the vulnerability and version levels. Specifically, vulnerability-level accuracy is typically lower than the No-Miss Ratio, and both are often lower than version-level F1-scores. 
This discrepancy stems from fundamental differences in evaluation granularity and tolerance to partial correctness. Vulnerability-level accuracy demands an exact match with the full set of affected versions, making it highly sensitive to both false positives and false negatives. The No-Miss Ratio relaxes this by tolerating over-estimation but still penalizes any missed affected version. In contrast, version-level metrics evaluate correctness at the granularity of individual versions, offering a more nuanced and forgiving view that better aligns with practical usage scenarios.
An exception is observed with \texttt{VULTURE}, where the NMR (55.1\%) exceeds the F1-score (42.9\%) due to its extremely high recall and low precision. This suggests that some tools prioritize broad coverage of potentially affected versions, achieving high recall and NMR at the cost of precision, thereby limiting their utility for precise vulnerability localization.

\begin{tcolorbox}[colback=black!10!white,enhanced,frame hidden, boxsep=0pt,left=5pt,right=5pt,top=2pt,bottom=2pt]
\finding
% Existing tools consistently perform better in approximating affected version ranges than in precisely identifying complete vulnerability scopes.
Existing tools cannot precisely identify the full affected versions of vulnerabilities (with no version omissions or false positives). However, they achieve an higher F1-scores in approximating the affected versions.

% \bella{This summarization may not be accurate. Could we say `Most existing tools tend to recall more affected versions, which yield a high recall while, in turn, sacrificing the precision.'?}
\end{tcolorbox}

\textbf{Methodology-Level Comparison.}
Tracing-based tools tend to outperform matching-based ones across both evaluation levels, yielding higher summary accuracy (38.4\% vs. 26.7\%), No-Miss Ratio (60.9\% vs. 44.6\%), and F1-score (72.9\% vs. 62.2\%). Most tracing-based tools (except \texttt{TC-SZZ}) exceed 40.0\% in accuracy and 55.0\% in No-Miss Ratio—thresholds that are typically not reached by matching-based tools, which often prioritize precision over completeness.
Version-level metrics reveal important exceptions. Matching-based tools such as \texttt{MOVERY} (72.8\%), \texttt{ReDeBug} (72.5\%), and \texttt{FIRE} (72.3\%) achieve F1-scores comparable to those of tracing-based tools like \texttt{LLM4SZZ} (72.3\%). In terms of precision, matching-based tools often achieve superior results—\texttt{VUDDY} and \texttt{V1SCAN} reach 94.2\% and 92.3\%, though typically at the cost of recall (e.g., \texttt{VUDDY}: 33.4\%). By contrast, tools like \texttt{VULTURE} prioritize recall (52.6\%) over precision (36.2\%), reflecting a design bias toward completeness rather than correctness.

These disparities reflect fundamental methodological differences. Matching-based tools compare code at the block/function level using syntactic or hash-based similarity, achieving high precision but remaining fragile to refactorings and minor edits. Tracing-based tools, by contrast, exploit fine-grained historical signals (e.g., \texttt{git blame}) to capture broader temporal context, offering better recall and resilience to superficial changes, yet potentially missing semantic dependencies beyond modified lines. We analyze these trade-offs further in RQ3.

\begin{tcolorbox}[colback=black!10!white,enhanced,frame hidden, boxsep=0pt,left=5pt,right=5pt,top=2pt,bottom=2pt]
\finding
While tracing-based tools outperform matching-based ones on average, several matching-based tools achieve comparable F1 scores at the version level and often outperform in precision.
\end{tcolorbox}

\subsection{Root Cause Analysis (RQ2)}
\label{rq3}

\begin{table*}[htbp]
  \centering
  \caption{Overview of the key stages and strategy variants adopted by tracing-based methods.}
  \resizebox{0.9\textwidth}{!}{%
  \renewcommand{\arraystretch}{0.6}
  \setlength{\tabcolsep}{1.3mm}
  \begin{tabular}{>{\centering\arraybackslash}m{10em}>{\centering\arraybackslash}m{16em}ccccccc}
    \toprule
    \textbf{Stage} &  \textbf{Strategy} & \textbf{VCCFinder} & \textbf{V-SZZ} & \textbf{Lifetime} & \textbf{SEM-SZZ} & \textbf{TC-SZZ}  & \textbf{LLM4SZZ}  \\
    \midrule
    \multirow{3}[6]{10em}{\centering \textbf{S1: Statements Selection}} 
    & Deleted or context lines(heuristic) & \ding{51}     & \ding{51}     & \ding{51}     &       &  \ding{51}       &  \\
    \cmidrule{2-9}
    & Data/control flow-based selection &       &       &       & \ding{51}     &        &  \\
    \cmidrule{2-9}
    & LLM-based selection &       &       &       &      &       & \ding{51} \\
    \cmidrule{1-9}          
    \multirow{3}[7]{10em}{\centering \textbf{S2: Commit Tracing}} 
    & Single-step blame tracing &   \ding{51}    &      & \ding{51}     &    \ding{51}   &       & \ding{51} \\
    \cmidrule{2-9}   
    & Iterative tracing with similarity filter &      &   \ding{51}    &       &      &         &  \\
    \cmidrule{2-9}   
    & Iterative tracing until the initial commit &      &      &       &      &  \ding{51}       &  \\
    \cmidrule{1-9}          
    \multirow{4}[5]{10em}{\centering \textbf{S3: Vulnerability-inducing Commit Selection}} 
    & Earliest candidate commit &      &   \ding{51}    &       &     &    \ding{51}     &  \\
    \cmidrule{2-9}          
    & Most-blamed commit &   \ding{51}    &      &   \ding{51}   &       &         &  \\
    \cmidrule{2-9}          
    & Commit covering all target lines &   &   &   & \ding{51} &    &   \\
    \cmidrule{2-9} 
    & LLM-selected commit &   &   &   &  &    & \ding{51}  \\
    \cmidrule{1-9}          
    \multirow{2}[1]{10em}{\centering \textbf{S4: Affected Versions Inference}} 
    & With cross-branch patch reuse &   & \ding{51} &       &       &          &  \\
    \cmidrule{2-9}          
    & Without cross-branch patch reuse & \ding{51} &      & \ding{51}     & \ding{51}     & \ding{51}        & \ding{51} \\
    \bottomrule
  \end{tabular}%
  }
  \label{tab:tracing_strategy_taxonomy}%
\end{table*}

 %tab:tracing_strategy_taxonomy
% Table generated by Excel2LaTeX from sheet 'Sheet4'
\begin{table*}[htbp]
  \centering
  \caption{Overview of the key stages and strategy variants adopted by matching-based methods.}
  \renewcommand{\arraystretch}{0.7}
  \setlength{\tabcolsep}{1.3mm}
  \resizebox{0.9\textwidth}{!}{
    \begin{tabular}{>{\centering\arraybackslash}m{10em}>{\centering\arraybackslash}m{20em}ccccccc}
    \toprule

\multicolumn{1}{c}{\textbf{Stage}} & \multicolumn{1}{c}{\textbf{Strategy}} & \textbf{ReDeBug} & \textbf{VUDDY}  & \textbf{MOVERY} & \textbf{V1SCAN} & \textbf{FIRE} & \textbf{VULTURE} \\
\midrule

\multirow{5}[8]{10em}{\centering \textbf{S1: Vulnerability Signature Construction}} 
    & Pre-patch Context Signature & \ding{51} &     &   &   &   &  \\
\cmidrule{2-9}
    & Modified statement signature &   &   &   \ding{51}    & \ding{51} & \ding{51} & \ding{51}  \\
\cmidrule{2-9}
    & Entire pre-patch function signature  &   & \ding{51}  & \ding{51}  & \ding{51} & \ding{51} & \ding{51} \\
\cmidrule{2-9}          
    & Post-patch function signature &   &    &   &  \ding{51} &  \ding{51}  &  \ding{51} \\
\cmidrule{2-9}          
    & Semantically related statements &   &      & \ding{51}  &   & \ding{51}  &  \\

\cmidrule{1-9}          

\multirow{4}[5]{10em}{\centering \textbf{S2: Matching}} 
    & Exact Matching & \ding{51}     & \ding{51}         & \ding{51}     & \ding{51}     & \ding{51}     & \ding{51} \\
\cmidrule{2-9}  
    & Approximate Matching via LSH &       &        &       &  \ding{51}     &      &  \ding{51}\\
\cmidrule{2-9}  
    & Approximate Matching via similarity metrics  &       &          &       &       & \ding{51}     &  \\
    \bottomrule
    \end{tabular}%
    }
\label{tab:matching_strategy_taxonomy}%
\end{table*}%

 %tab:matching_strategy_taxonomy

\subsubsection{\textbf{Setup}}
To understand the root causes of FPs and FNs in vulnerability-affected version identification, we adopt a mixed-method approach combining qualitative analysis and quantitative validation. We first systematically examine the key technical strategies used in tracing-based and matching-based methods, analyzing potential limitations at each stage. To validate these observations, we randomly sample 100 vulnerabilities from our dataset and evaluate identification results under representative strategies.

\subsubsection{\textbf{Analysis of Tracing-based Tools}}
Table~\ref{tab:tracing_strategy_taxonomy} summarizes the strategies adopted by representative tracing-based tools across four critical stages: \textit{Statement Selection}, \textit{Commit Tracing}, \textit{Vulnerability-Inducing Commit Identification}, and \textit{Affected Version Inference}.

\textit{\textbf{a) Statement Selection:}}
Most tracing-based tools (e.g., \texttt{V-SZZ}, \texttt{VCCFinder}, \texttt{Lifetime}, \texttt{TC-SZZ}) rely on simple patch-based heuristics to select tracing targets, typically focusing on deleted lines or those adjacent to additions. \texttt{V-SZZ} and \texttt{TC-SZZ} are even limited to patches with deletions. While these heuristics are straightforward and easy to implement, they often fail to capture the semantic core of the vulnerability.
Our manual analysis of 100 representative vulnerabilities reveals two primary sources of error in this selection process.
First, when patches span multiple functions or files, most existing tools treat all code hunks uniformly, without distinguishing between semantically relevant and irrelevant edits. The lack of granularity results in the inclusion of noisy, non-critical changes as tracing candidates, which weakens the signal used for version identification. Among the analyzed cases, 49 patches involve modifications across multiple functions or files, of which 16 contain such irrelevant hunks. Although \texttt{LLM4SZZ} try to address this issue using LLM for contextual filtering, it correctly excluded the noise in only 9 of these 16 cases—indicating limited success even with advanced semantic modeling.

Second, even in patches that modify only a single function, heuristic-based methods frequently fail to select the correct vulnerable statements. It is mainly due to their reliance on surface-level syntactic proximity rather than any principled understanding of the vulnerability’s root cause. Consequently, statement selection remains brittle in the absence of deeper semantic analysis.
Across the full set of 100 cases, these limitations led to incorrect statement selection in 49 instances. More sophisticated techniques such as semantic dependency analysis (\texttt{SEM-SZZ}) and LLM-based inference (\texttt{LLM4SZZ}) have been proposed to mitigate this issue. Nonetheless, these methods still incorrectly identified relevant code in 39 and 28 cases, respectively, underscoring persistent challenges in semantic reasoning and contextual disambiguation.

% Most tools (e.g., \texttt{V-SZZ}, \texttt{VCCFinder}, \texttt{Lifetime}, \texttt{TC-SZZ}) adopt simple patch-based heuristics that select deleted lines or lines adjacent to additions as tracing targets. \texttt{V-SZZ} is even limited to patches with deletions. While easy to implement, these heuristics often fail to capture the semantic essence of the vulnerability. In our manual study, this strategy led to incorrect statement selection in 49 vulnerability cases.
% To improve upon this, \texttt{SEM-SZZ} integrates semantic dependency analysis, and \texttt{LLM4SZZ} uses large language models (LLMs) to identify vulnerable statements. However, these tools still misidentified the relevant code in 39 and 28 cases respectively, highlighting limitations in semantic reasoning and context modeling.

\textit{\textbf{b) Commit Tracing:}}
Most tools (e.g., \texttt{VCCFinder}, \texttt{Lifetime}, \texttt{SEM-SZZ}, \texttt{LLM4SZZ}) apply one-step backward tracing, which often misses earlier commits that introduced the vulnerability. Among the 100 cases, only 70 were traceable via a single step, whereas 30 required multi-step tracing. The strategy of \texttt{TC-SZZ} tracing back to the initial commit, however, led to more false positives, with version-level precision being only 59.3\%, which significantly lower than \texttt{V-SZZ}'s 74.5\%. Iterative tracing methods used in \texttt{V-SZZ} attempt to address this via similarity-based heuristics but still suffer from inaccuracies: 16 were over-traced and 12 were under-traced.

\textit{\textbf{c) Vulnerability-Inducing Commit Identification:}}
Once candidate commits are obtained, tools adopt varying heuristics to identify the final vulnerability-inducing commit, such as selecting the earliest or most frequently blamed commit. These heuristics, however, do not verify whether the selected commit actually introduced the vulnerability. \texttt{LLM4SZZ} try semantic verification by prompting an LLM to check vulnerability existence prior to a given commit. Nevertheless, our evaluation on 100 cases showed that \texttt{LLM4SZZ} still produced \textbf{12} FPs and \textbf{29} FNs, suggesting the difficulty of reliable semantic reasoning at the commit level.

\textit{\textbf{d) Affected Version Inference:}}
In multi-branch development environments, patches may be inconsistently propagated. Tools that only analyze the \texttt{main} branch (e.g., \texttt{V-SZZ}) risk missing relevant versions. Though cross-branch matching is supported, \texttt{V-SZZ}'s simple criteria for finding duplicated patches still failed to detect relevant patches in 13 out of 100 cases.

\begin{tcolorbox}[colback=black!10!white,enhanced,frame hidden, boxsep=0pt,left=5pt,right=5pt,top=2pt,bottom=2pt]
\finding
Tracing-based approaches are sensitive to the accuracy of identified vulnerability statements. While semantic analysis and LLMs enhance performance, key limitations persist. Moreover, their effectiveness is further constrained by the inaccuracy in selecting vulnerability-inducing commits and the strategy of commit tracing.
% Tracing-based methods are constrained by static heuristics, resulting in substantial FPs and FNs. While semantic analysis and LLMs enhance performance, key limitations persist.
\end{tcolorbox}

% In summary, tracing-based tools heavily rely on static heuristics at every stage. Their inability to robustly model code semantics, temporal evolution, and branching strategies in real-world repositories leads to frequent FPs and FNs.

\subsubsection{\textbf{Analysis of Matching-based Tools}}
Table~\ref{tab:matching_strategy_taxonomy} summarizes the taxonomy of matching-based tools, which typically follow two phases: \textit{vulnerability signature construction} and \textit{signature matching}.

\textit{\textbf{a) Vulnerability Signature Construction:}}
Tools in this category extract syntactic or semantic features from known vulnerability patches to match similar instances in other versions.
However, the granularity and flexibility of the extracted signatures significantly affect their effectiveness.
Early tools such as \texttt{ReDebug} and \texttt{VUDDY} employ coarse-grained strategies: \texttt{ReDebug} builds sliding-window token sequences over patch hunks, while \texttt{VUDDY} uses the entire pre-patch function as a matching signature. These coarse-grained strategies are sensitive to unrelated code edits, resulting in frequent FNs.
To improve specificity, \texttt{Movery} and \texttt{FIRE} extract vulnerability-relevant statements via semantic analysis. However, both tools apply uniform extraction rules regardless of patch semantics or vulnerability type. For example, \texttt{Movery} incorrectly extracted features in 47 cases revealing limitations in generalized semantic modeling.

Moreover, tools including \texttt{Movery}, \texttt{V1SCAN}, \texttt{FIRE}, and \texttt{VULTURE} further assume that deleted lines in the patch must be present in the target code for a match. This rigid assumption fails to accommodate variants due to refactoring or non-deletion-based repairs. In our dataset, this assumption led to 55.8\% of deletable-line vulnerabilities being overlooked.
Notably, these tools share a structural weakness with tracing-based methods: they treat all patch modifications as equally relevant. For patches spanning multiple functions or files, this strategy often leads to semantically unrelated edits being included in the extracted signatures, thereby increasing FPs.

\textit{\textbf{b) Signature Matching:}}
Most tools perform \textit{exact matching} between extracted features and target code, which is brittle and contributes to high FN rates. While tools (e.g., \texttt{V1SCAN}, \texttt{FIRE}, \texttt{VULTURE}) relax matching using structural or semantic similarity, none verify whether the matched code actually contains the vulnerability. Consequently, these tools either fail to identify semantically equivalent vulnerabilities or misidentify safe code as vulnerable.
FPs also stem from flawed signature granularity. \texttt{VUDDY}'s function-level signatures, for instance, produce incorrect matches in two scenarios: (1) lack of structural normalization—if the patch modifies a function’s structure (e.g., through reordering or renaming), matching may erroneously flag non-vulnerable code; (2) structural duplicates—older versions may contain functions with similar structure but semantically unrelated functions, which are wrongly matched due to lack of inter-procedural reasoning.

\begin{tcolorbox}[colback=black!10!white,enhanced,frame hidden, boxsep=0pt,left=5pt,right=5pt,top=2pt,bottom=2pt]
\finding
The use of coarse-grained features, inflexible matching strategies, and insufficient semantic modeling limits the ability of matching-based methods to accurately identify vulnerability-affected versions.
\end{tcolorbox}

\subsection{Patch-Type Sensitivity Analysis (RQ3) }
\label{rq2}

% \begin{figure}
%     \centering
%     \includegraphics[width=\linewidth]{fig/diff_patch_bar_cropped.pdf}
%     \caption{Impact of Patch Modification Types on Tool Performance (Add-only, Del-only, Mixed). }
%     \label{fig:diff_patch_bar}
% \end{figure}

% \input{table/only_added&added_deled_lines}

% \begin{figure}
%     \centering
%     \includegraphics[width=\linewidth]{fig/multi_func_bar_cropped.pdf}
%     \caption{Tool Performance under Varying Patch Modification Scopes. SF: Single function; MFSF: Multiple functions within a single file; MF: Multiple files.}
%     \label{fig:multi_function_bar}
% \end{figure}

\begin{figure*}[htbp] 
    \centering 

    \subfloat[Added-only]{%
        \includegraphics[width=0.325\textwidth]{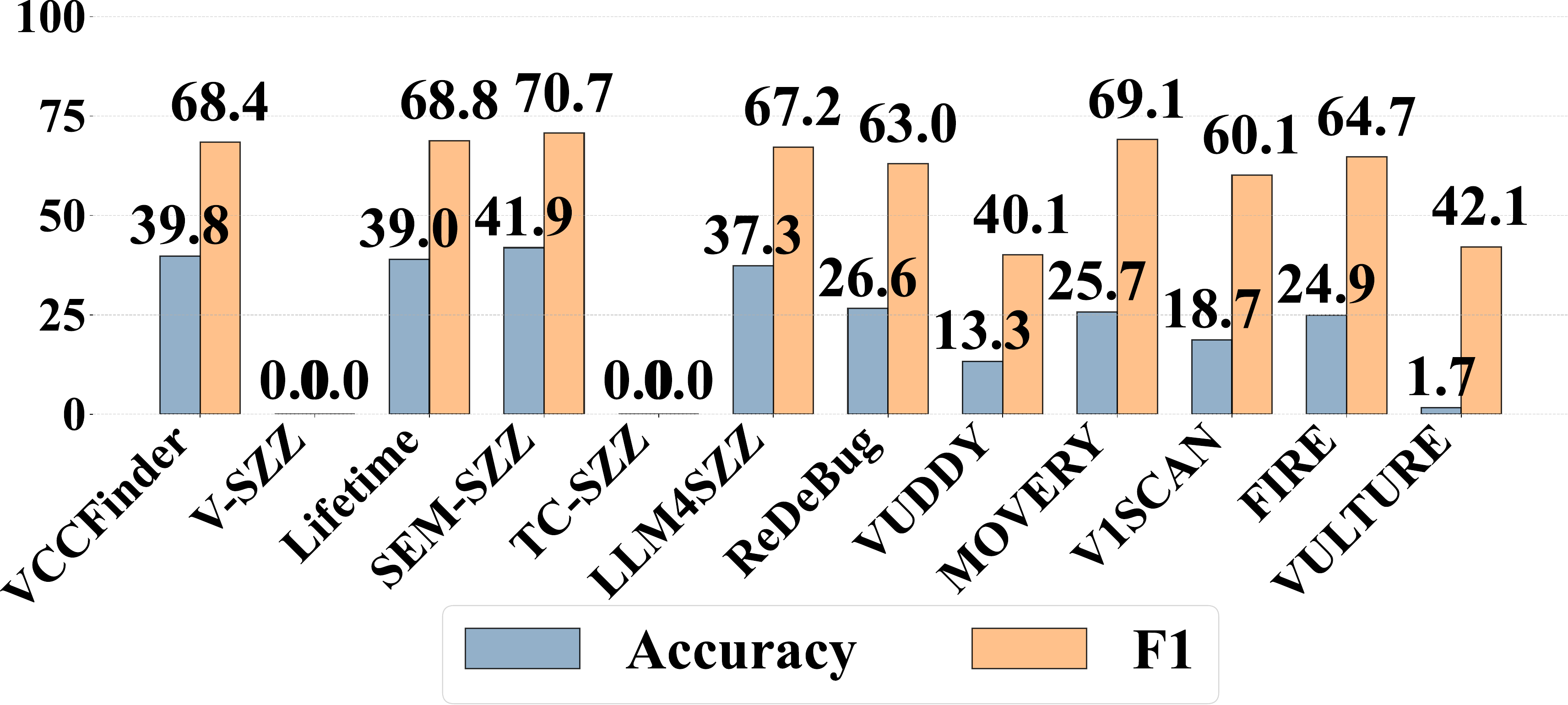}%
        \label{fig:sub_addonly}
    }
    \hfill
    \subfloat[Del-only]{%
        \includegraphics[width=0.325\textwidth]{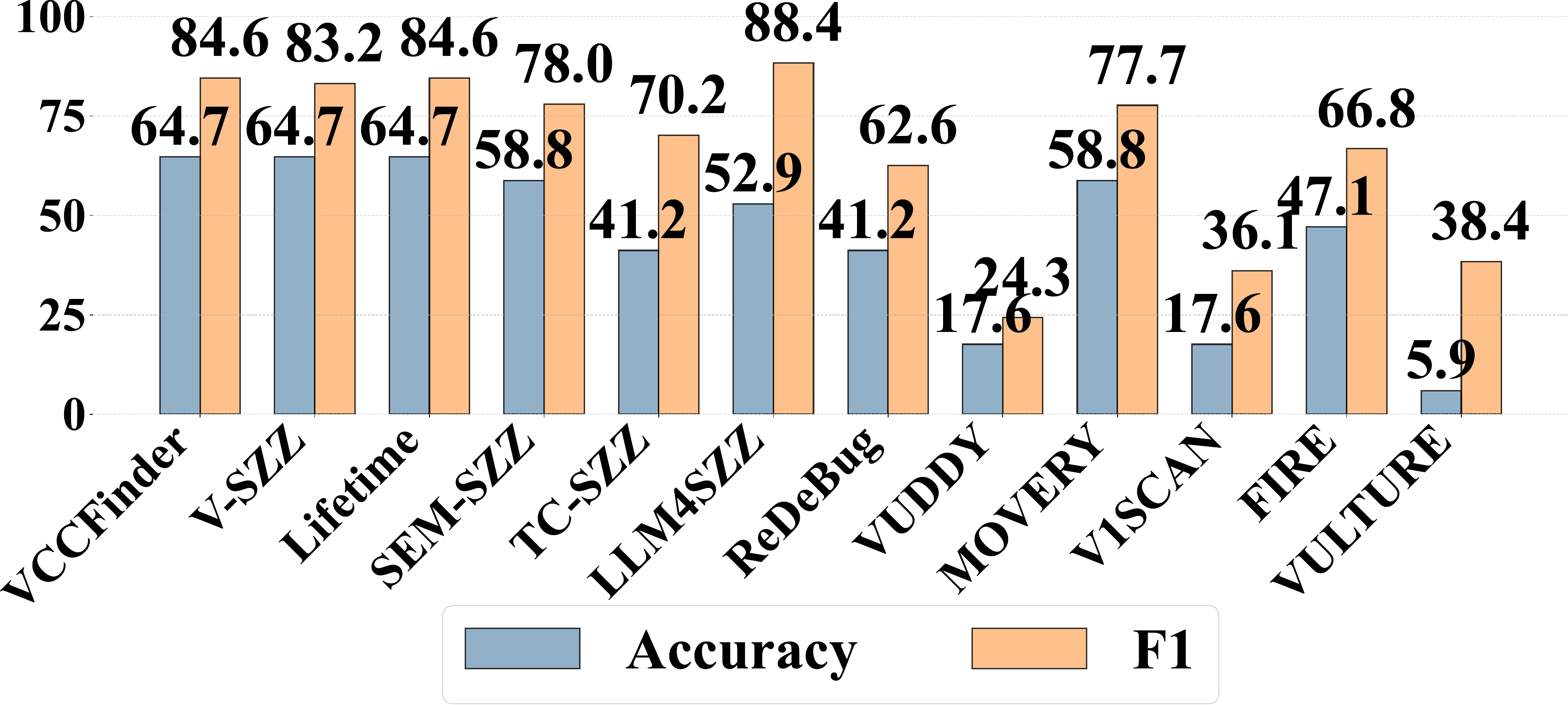}%
        \label{fig:sub_delonly}
    }
    \hfill
    \subfloat[Mixed Patches]{%
        \includegraphics[width=0.325\textwidth]{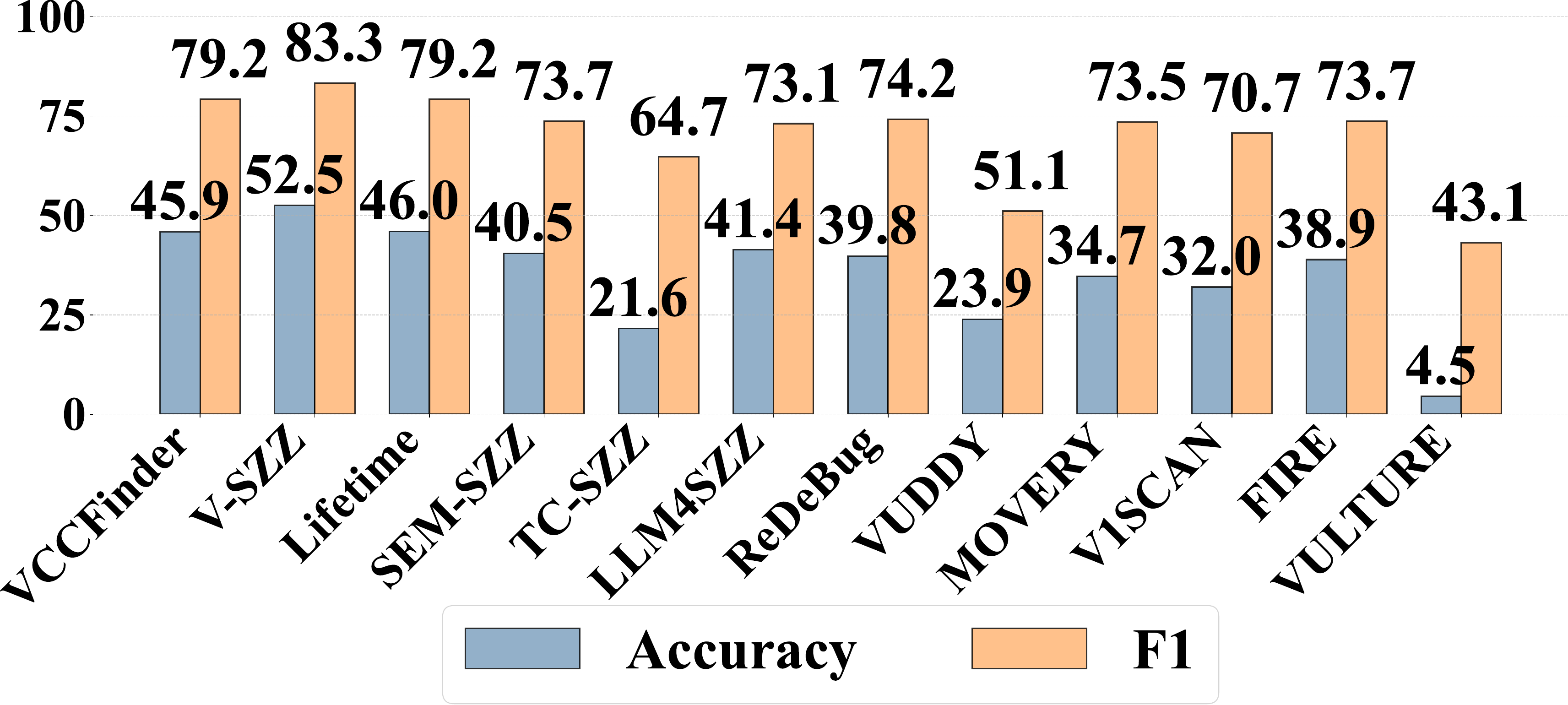}%
        \label{fig:sub_mixed}
    }

    \caption{Impact of Patch Modification Types on Tool
    Performance (Add-only, Del-only, Mixed).} 
    \label{fig:patch_modification} 
\end{figure*}

\begin{figure*}[!htbp] 
    \centering 

    \subfloat[Single-Function\label{fig:sub_singlefunc}]{
        \includegraphics[width=0.31\textwidth]{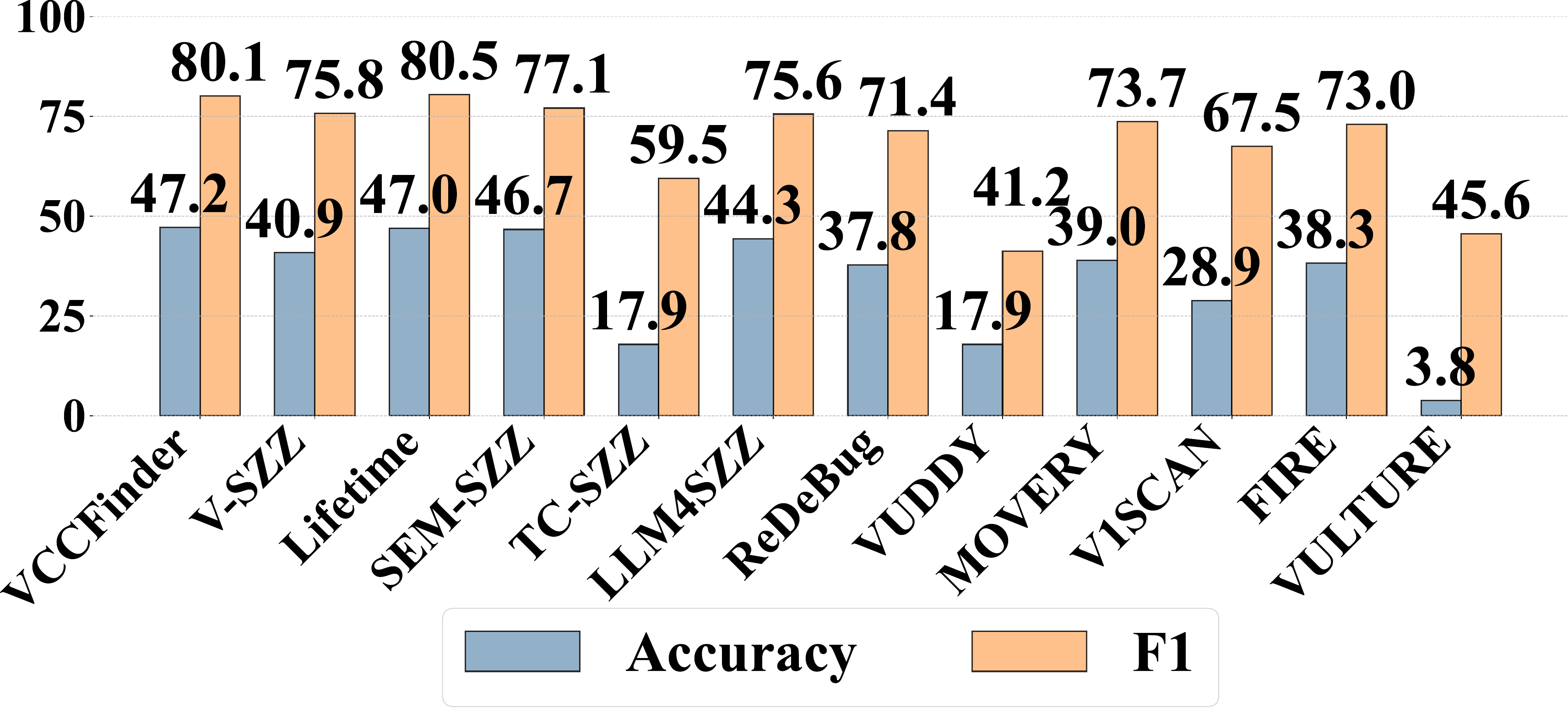}
    }
    \hfill
    \subfloat[Multi-Function Single-File\label{fig:sub_mutlifuncsinglefile}]{
        \includegraphics[width=0.31\textwidth]{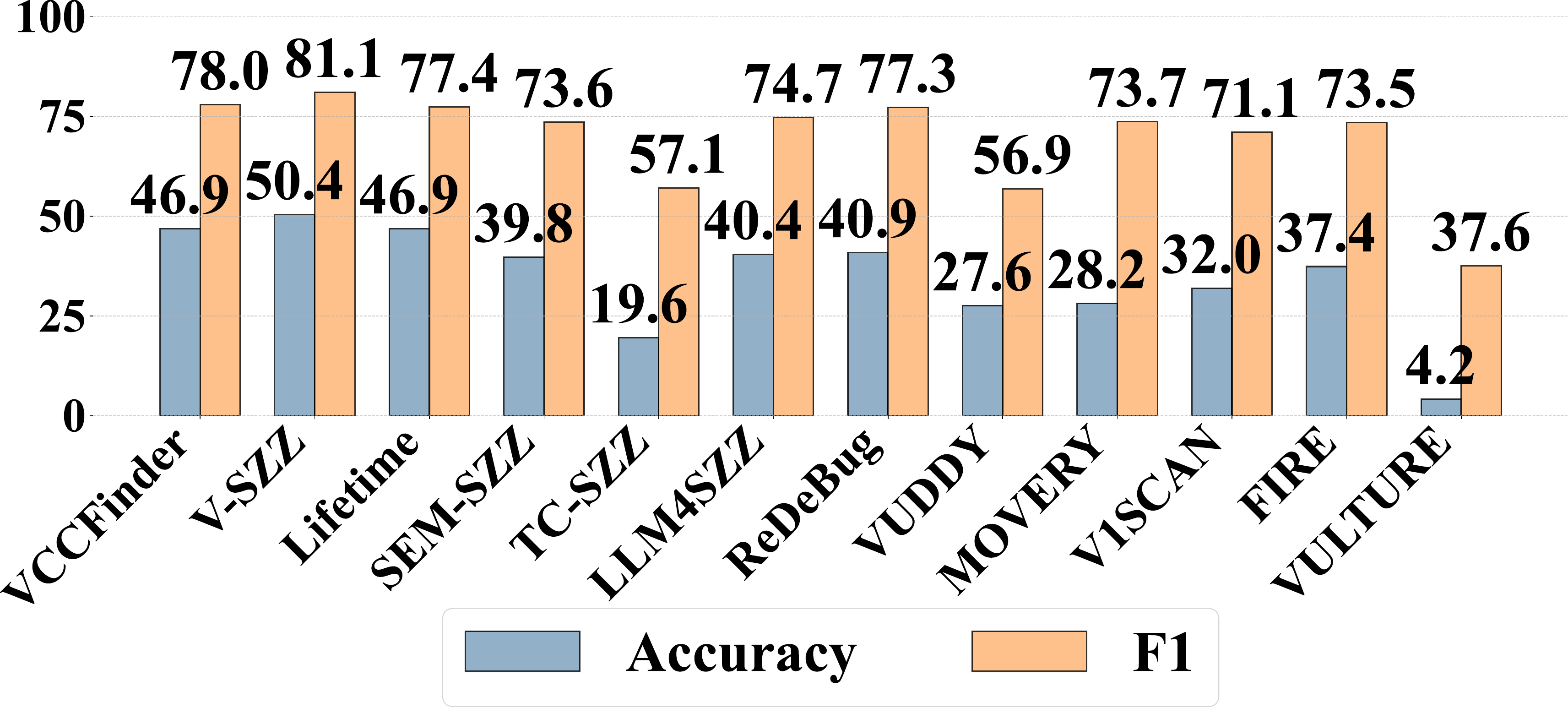}
    }
    \hfill
    \subfloat[Multi-Function Multi-File\label{fig:sub_mutlifuncmultifile}]{
        \includegraphics[width=0.31\textwidth]{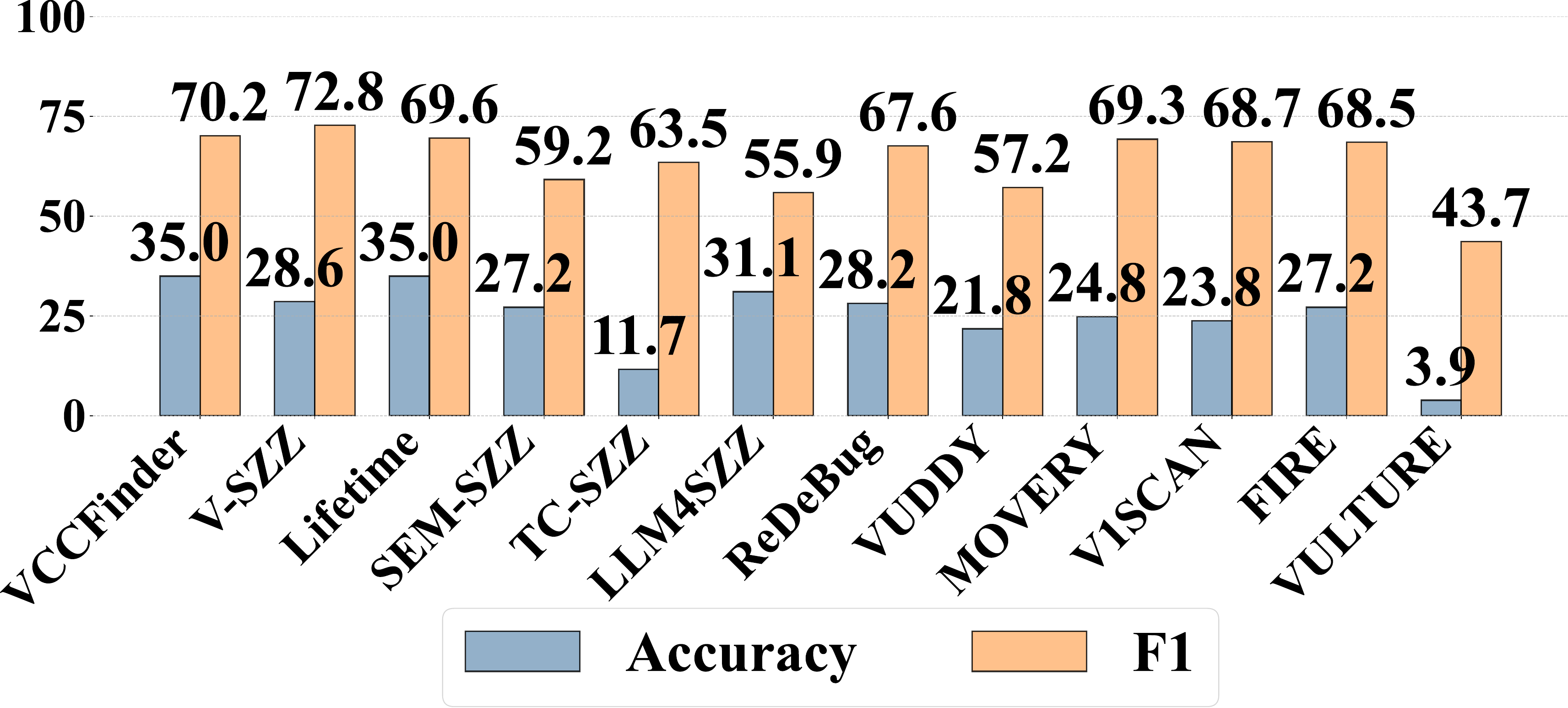}
    }

    \caption{Tool Performance under Varying Patch Modification Scopes.} 
    \label{fig:function_performance} 
\end{figure*}

While RQ1 and RQ2 assess tool effectiveness from overall and stage-specific perspectives, they leave open the question of how structural properties of vulnerability patches influence performance. Thus, we analyze tool robustness across three patch-level dimensions, based on their direct alignment with the internal assumptions commonly made by existing tracing- and matching-based methods. The insights also support the design and evaluation of ensemble configurations in RQ4.

\begin{itemize}[leftmargin=*]
    \item \textbf{Type of Code Changes.} As observed in RQ2, most tools rely heavily on deleted lines for tracing. We therefore classify patches into \emph{Add-only}, \emph{Del-only}, and \emph{Mixed} types, to evaluate sensitivity of this structural dependency.

    \item \textbf{Scope of Modifications.} Tools typically treat all modifications uniformly, regardless of how code changes are localized. However, real-world patches may cantain a single function, span multiple functions within a file, or modify code across files. This dimension helps assess each tool’s resilience to dispersed or concentrated changes.

    \item \textbf{Cross-Branch Context.} In multi-branch development workflows, the same vulnerability may be patched differently across branches. Although \texttt{V-SZZ} incorporates cross-branch information when identifying affected versions, the general impact of single-branch versus multi-branch patches on tool performance has not been fully explored/determined.
    While \texttt{V-SZZ} incorporates cross-branch information when identifying affected versions, the general impact of single-branch versus multi-branch patches on tool performance remains underexplored.
\end{itemize}

\subsubsection{\textbf{Setup}}
Based on the above design, we evaluate tools along with these three patch dimensions to understand their structural robustness and uncover potential blind spots.

% To understand the robustness and generalization of existing tools, we analyze their performance across three key dimensions:

% \begin{itemize}
%     \item \textbf{Patch characteristics.} We consider (1) the type of code changes—patches with only additions, only deletions, or both; and (2) the scope of changes—modifications within a single function, a single file, or across multiple files. These factors reflect the structural complexity of patches, which may challenge tracing- or matching-based methods differently.
    
%     \item \textbf{Vulnerability type.} We group CVEs based on their CWE categories (e.g., buffer overflows, null pointer deference) to investigate whether tool performance varies across different vulnerability classes.
    
%     \item \textbf{Project maintenance model.} We distinguish between projects with a single active branch and those with long-term multi-branch maintenance, as branching may complicate the version history and affect detection accuracy.
% \end{itemize}

\subsubsection{\textbf{Results}}

\textbf{Impact of Type of Code Changes: Add-only, Del-only, Mixed.}
\figu~\ref{fig:patch_modification} shows tool performance across three patch types.
Tracing-based tools perform well on \emph{Del-only} and \emph{Mixed} patches, but their performance drops sharply on \emph{Add-only} patches, with version-level F1 decreasing by over 10 percentage points. This is because these tools rely on deleted lines to initiate blame tracing. Without deletions, tools like \texttt{V-SZZ} and \texttt{TC-SZZ} cannot operate. \texttt{SEM-SZZ} is more robust, as its dependency analysis allows itself to extract signals beyond simple line deletions.
This observation
% \bella{Which pattern? Do you mean `patch'? I get a bit lost here. Why this observation can explain the semantics?} 
reflects the semantics of patch content: deleted lines often contain the root cause, while added lines typically represent mitigation and offer limited cues for identifying the original vulnerability. Without deletions, tracing tools lack meaningful anchors to begin backward analysis.

Matching-based tools follow a different trend. They perform best on \emph{Mixed} patches, where both added and deleted lines provide richer context for constructing semantic signatures. Performance declines on \emph{Add-only} and \emph{Del-only} patches due to reduced context. In particular, \emph{Add-only} patches are more challenging, as they contain mitigation logic rather than faulty code, making semantic matching less effective.

Performance also varies by tool: \texttt{SEM-SZZ} performs best on \emph{Add-only} patches, \texttt{LLM4SZZ} on \emph{Del-only}, and \texttt{V-SZZ} on \emph{Mixed} patches. This diversity highlights the complementarity of different approaches and motivates the ensemble configurations explored in RQ4.

\emph{Del-only} patches are rare (1.3\%), while \emph{Add-only} and \emph{Mixed} patches account for 21.3\% and 77.4\% of the dataset, respectively (Table~\ref{tab:dataset_details}). Improving robustness on these dominant types—especially \emph{Add-only}—is therefore essential for real-world applicability.

\begin{tcolorbox}[colback=black!10!white,enhanced,frame hidden, boxsep=0pt,left=5pt,right=5pt,top=2pt,bottom=2pt]
\finding
Tracing-based tools fail on \emph{Add-only} patches due to their reliance on deletions. Matching-based tools perform better on \emph{Mixed} patches, where both additions and deletions provide richer context.
\end{tcolorbox}

\textbf{Impact of Scope of Modification: Function-, File-, and Multi-File Changes.}
\figu~\ref{fig:function_performance} compares tool performance across different patch scopes. Both tracing-based and matching-based methods perform similarly on patches confined to a single function or multiple functions within the same file. However, their effectiveness drops significantly on \emph{multi-file} patches.
A closer inspection reveals that multi-file patches introduce substantially more non-vulnerability-related changes. In a sample of 50 patches from each category, multi-file patches (20) included significantly more non-vulnerability-related changes than single-file ones (7), adding noise and hindering accurate detection.

Among all tools, recall tends to increase while precision declines as the patch scope expands. This is due to coarse-grained analysis: tracing-based tools operate at the hunk level, while matching-based tools assess at the function level. A single matched function may cause an entire patch to be flagged, increasing recall but reducing precision.

In terms of performance, \texttt{Lifetime} achieves the highest F1 on \emph{single-function} patches, while \texttt{V-SZZ} performs the best on both \emph{single-file} and \emph{multi-file} patches. The dataset includes 585 single-function (51.9\%), 337 single-file (29.9\%), and 206 multi-file patches (18.2\%). Though multi-file patches are less frequent, they still constitute a substantial portion of real-world cases and remain a key challenge.

% We found these tools performed poorly in patches with multi-functions. We divided the patches into three types: single-function, multi-function single-file, multi-function multi-file. We observed that these tools except VUDDY have reduced F1 scores by more than 12\% in patches with multi-function in different files in \tabl~\ref{tab:diff_func_result}. Specifically, V-SZZ only get A-F1 with 49.1\% in patches with multi-function multi-file. The main reason is that patches with multi-function are complex and the vulnerability statements often exist in different functions. These tools have not designed module for inter-procedural analysis, leading to poorly handling of these patches. If the multi-function patches simply split into isolated function, these dependency relationships are ignored. So, this is crucial to employ inter-procedural analysis within those functions to generate accurate vulnerability statements.

\begin{tcolorbox}[colback=black!10!white,enhanced,frame hidden, boxsep=0pt,left=5pt,right=5pt,top=2pt,bottom=2pt]
\finding
Multi-file patches introduce more irrelevant changes, leading to notable performance drops across tools. Although less common, they pose a significant challenge and warrant focused attention.
\end{tcolorbox}

\begin{figure}[htbp!] 
    \centering 

    \subfloat[Single-Branch\label{fig:sub_single_branch}]{
        \includegraphics[width=0.47\columnwidth]{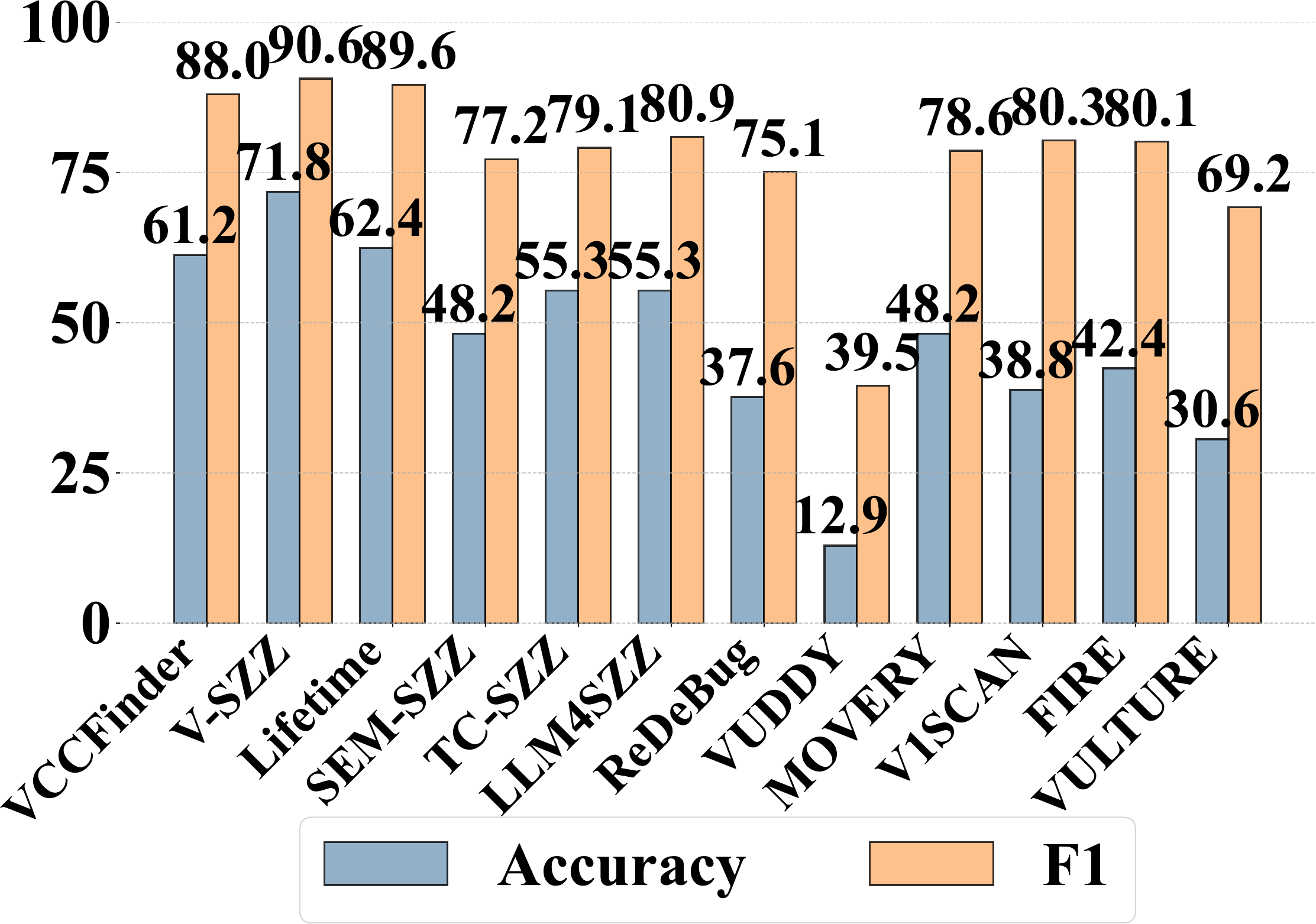}
    }
    \hfill
    \subfloat[Multi-Branch\label{fig:sub_multi_branch}]{
        \includegraphics[width=0.47\columnwidth]{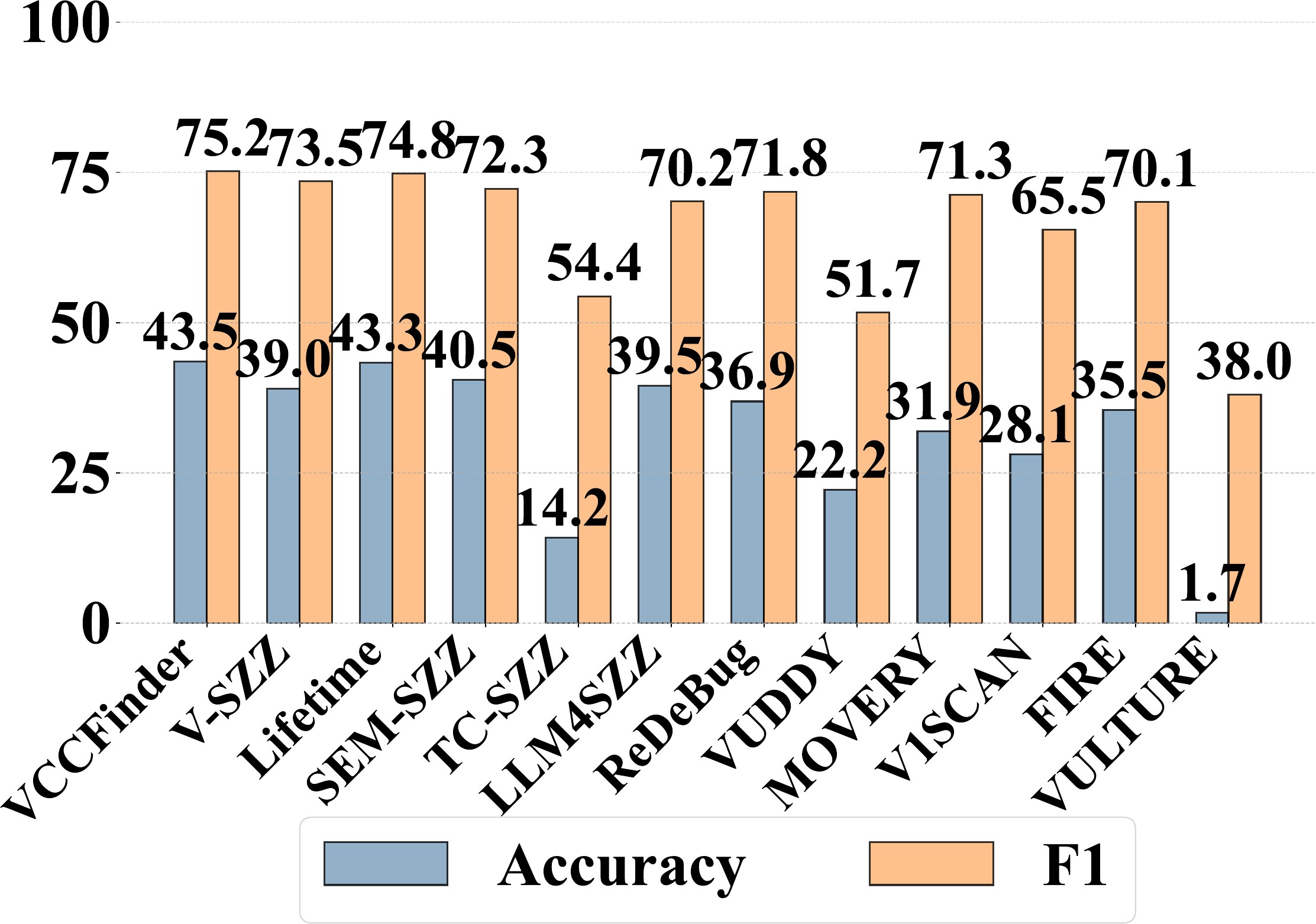}
    }

    \caption{Performance of Different Tools under Single-Branch and Multi-Branch Projects.} 
    \label{fig:branch_performance}
\end{figure}

\textbf{Impact of Cross-Branch: Single-branch vs Multi-branch.}
\figu~\ref{fig:branch_performance} summarizes tool performance under single-branch and multi-branch settings. In multi-branch scenarios most tools suffer marked performance declines, for instance, \texttt{VULTURE}’s F1-score plunges by 31.2\% and \texttt{V-SZZ}’s accuracy drops 32.7\%, highlighting the added complexity of identifying affected versions across diverging codebases.
% \bella{Can we add statistics here?}
Among tracing-based tools, \texttt{V-SZZ} is the only one that incorporates cross-branch information during version inference. This strategy helps mitigate recall degradation in multi-branch settings. However, its multi-step tracing introduces noise from unrelated commits, which reduces precision and partially offsets the benefit. Other tracing tools rely solely on patches from the main branch and thus miss cross-branch vulnerabilities, leading to larger performance declines.
Matching-based tools are more severely affected by multi-branch development due to increased structural variability across branches. Prior work~\cite{Yang2023EnhancingOP} shows that only 9.37\% of patches are syntactically identical across branches, limiting the effectiveness of exact or near-exact matching strategies used by tools such as \texttt{MOVERY} and \texttt{FIRE}.
Overall, identifying affected versions in multi-branch projects remains a challenging task.

% Finally, we divided the dataset into single-branch and multi-branch categories based on the collaborative development of the open-source projects to conduct a comparative analysis. \tabl~\ref{tab:result of col development} shows that for both tracing-based and matching-based mthods, accuracy in multi-branch projects is significantly lower than in single-branch projects. This degradation can be attributed to two main factors. First, in multi-branch development, security patches are often propagated via cross-branch merges, yet existing detection tools (with the exception of V-SZZ) generally lack a mechanism for tracing patches across branches. That leads tracing-based methods to misidentify patched versions. Second, multi-branch projects typically involve more complex collaboration patterns, they usually perform more refactoring or modifications on the code than single-branch projects. This diversity of code changes greatly complicates the matching of vulnerability features of match-based methods. 

\begin{tcolorbox}[colback=black!10!white,enhanced,frame hidden, boxsep=0pt,left=5pt,right=5pt,top=2pt,bottom=2pt]
\finding
Multi-branch development significantly degrades the effectiveness of both tracing-based and matching-based tools, due to limited cross-branch patch utilization and substantial code divergence across branches.
\end{tcolorbox}

\subsection{Tool Combination Analysis (RQ4)}
\label{rq4}

Based on earlier findings, this RQ investigates whether combining tool components or outputs can yield performance improvements over standalone tools. We explore this from two angles: modular recomposition of tracing-based tools and ensemble strategies spanning tracing- and matching-based tools.

\subsubsection{\textbf{Setup}}
Our evaluation is organized into two phases:

\textbf{Phase-1: Modular Recomposition of Tracing-based Tools.}
We focus on tracing-based tools because their modular workflows are amenable to recombination, whereas the core stages of matching-based tools are often tightly coupled and less separable. Based on the stage-wise decomposition in RQ2, we identify four stages: (S1) statement selection, (S2) impact range inference, (S3) commit tracing, and (S4) cross-branch patch reuse. Prior results show that LLM-based methods in S1 and patch propagation in S4 consistently outperform alternatives. Fixing these two stages, we systematically explore combinations of 2 alternatives in S2 and 4 in S3, yielding 7 hybrid configurations ($2 \times 4 - 1 = 7$). The best variant (\texttt{LLM4SZZ+}) serves as a representative for Phase-2.

\textbf{Phase-2: Cross-Tool Combination.}
This phase investigates whether outputs from diverse tools—spanning both tracing- and matching-based paradigms—can be effectively integrated. We select ten high-performing tools (F1-score $\geq$ 70\%): \texttt{VCCFinder}, \texttt{V-SZZ}, \texttt{Lifetime}, \texttt{SEM-SZZ}, \texttt{LLM4SZZ}, \texttt{ReDebug}, \texttt{Movery}, \texttt{FIRE}, \texttt{V1SCAN}, and the newly derived \texttt{LLM4SZZ+}. We evaluate three ensemble strategies as follows:

\begin{itemize}[leftmargin=*]
  \item \textbf{Inclusion Strategy.} The cumulative effect of integrating tools is evaluated by testing sizes 2 to 10 union sets.
  \item \textbf{Voting Strategy.} To assess consensus-based robustness, we evaluate all combinations of 3, 5, 7, and 9 tools, marking a version as affected if the majority agrees.
  \item \textbf{Best-in-Dimension Strategy.}  We also select the best-performing tool in each of four key dimensions identified in RQ3 (e.g., patch modeling, commit tracing) and aggregate their outputs, leveraging their complementary strengths.
\end{itemize}

% We evaluate whether integrating outputs from diverse tools—including \texttt{LLM4SZZ+}—can improve robustness and accuracy. 

% Three representative strategies are tested:

% \begin{itemize}
%     \item \textbf{Inclusion Strategy.} 

%     We select ten high-performing tools (each with a version-level F1-score above 70.0\%) spanning both tracing-based and matching-based approaches: \texttt{VCCFinder}, \texttt{V-SZZ}, \texttt{Lifetime}, \texttt{SEM-SZZ}, \texttt{LLM4SZZ}, \texttt{ReDebug}, \texttt{Movery}, \texttt{FIRE}, \texttt{V1SCAN}, and \texttt{LLM4SZZ+}. To assess the cumulative benefits of tool integration, we evaluate all possible union sets formed by selecting between 2 and 10 tools.
%     \item \textbf{Voting Strategy.} To explore the impact of consensus-based decision-making, we exhaustively test all combinations of 3, 5, 7, and 9 tools from the selected pool. A version is labeled as affected if more than half of the tools in a combination agree. This strategy aims to improve robustness by mitigating individual tool errors through collective agreement.
%     \item \textbf{Best-in-Dimension Strategy.} Guided by the results in RQ2, we identify the best-performing tool in each of four key dimensions and aggregate their outputs. This strategy evaluates whether leveraging the specialized strengths of different tools in targeted scenarios can deliver strong overall performance.

% \end{itemize}

\begin{table}[t!]
  \centering
  \caption{Performance of Representative Tool Combinations under Three Strategies.}
  \label{tab:combination_result}
  \setlength{\tabcolsep}{1.8mm}
  \renewcommand{\arraystretch}{1.0}
  \resizebox{0.9\linewidth}{!}{
  \begin{tabular}{>{\centering\arraybackslash}m{1.3cm} >{\centering\arraybackslash}m{3.8cm} >{\centering\arraybackslash}m{1.25cm} >{\centering\arraybackslash}m{1.3cm}}
    \toprule
    \textbf{Strategy} &  \textbf{Tool Combination} & \textbf{Vuln. Acc.} & \textbf{Version F1} \\
    \midrule
    \multirow{2}{*}{\textbf{Inclusion}} 
      & ReDeBug, LLM4SZZ+ & 52.1\% & 84.0\% \\
      & ReDeBug, V1SCAN, LLM4SZZ+ & 51.7\% & 84.5\% \\
    \midrule
    \textbf{Voting} 
      & V-SZZ, Lifetime, SEM-SZZ, Movery, LLM4SZZ+ & \textbf{55.0\%} & \textbf{84.8\%} \\
    \midrule
    \textbf{Best-in-Dimension} 
      & SEM-SZZ, LLM4SZZ, V-SZZ & 50.3\% & 81.7\% \\
    \midrule
    \multicolumn{2}{l}{\textit{Reference: Best Individual Tools}} & & \\
    \multicolumn{2}{l}{\quad VCCFinder (standalone)} & 44.9\% & 77.8\% \\
    \multicolumn{2}{l}{\quad LLM4SZZ+ (modular hybrid)} & 50.7\% & 81.8\% \\
    \bottomrule
  \end{tabular}
  }
\end{table}

\subsubsection{\textbf{Results}} 
Table~\ref{tab:combination_result} summarizes the top-performing configurations under each combination strategy.

\textbf{Modular Recomposition.}
The best configuration, \texttt{LLM4SZZ+}, integrates LLM-based statement selection (S1), single-step blame tracing (S2), full-line coverage for commit tracing (S3), and cross-branch patch reuse (S4). This variant achieves 50.7\% accuracy at the vulnerability level and 81.8\% F1-score at the version level—outperforming the strongest tracing-based baseline (\texttt{VCCFinder}) by 5.8\% and 4.0\%, and the earlier hybrid version (\texttt{LLM4SZZ}) by 10.0\% and 9.5\%.
Notably, replacing S3 component with LLM-selected commits significantly reduces performance (-7.3\% accuracy, -5.7\% F1), suggesting that though LLMs enhance early-stage selection, heuristic-based commit tracing remains more effective.

\begin{tcolorbox}[colback=black!10!white,enhanced,frame hidden, boxsep=0pt,left=5pt,right=5pt,top=2pt,bottom=2pt]
\finding
Modular recomposition yields great improvement. LLMs are best at 
% most effective for 
statement selection, whereas commit identification benefits more from heuristic methods.
\end{tcolorbox}

\textbf{Inclusion Strategy.}
The combination of \texttt{ReDebug} and \texttt{LLM4SZZ+} achieves the highest accuracy at the vulnerability level (52.1\%), an improvement of 7.2\% over \texttt{VCCFinder} and 1.4\% over \texttt{LLM4SZZ+}. At the version level, it reaches an F1-score of 84.0\%.
Another configuration—\texttt{ReDebug}, \texttt{V1SCAN}, and \texttt{LLM4SZZ+}—achieves the highest version-level F1-score (84.5\%), outperforming \texttt{VCCFinder} and \texttt{LLM4SZZ+} by 6.7\% and 2.7\%, with a vulnerability-level accuracy of 51.7\%.

\textbf{Voting Strategy.}
The optimal combination includes \texttt{V-SZZ}, \texttt{Lifetime}, \texttt{SEM-SZZ}, \texttt{Movery}, and \texttt{LLM4SZZ+}. This ensemble achieves 55.0\% accuracy at the vulnerability level and 84.8\% F1-score at the version level—improvements of 10.1\% and 7.0\% over \texttt{VCCFinder}, and 4.3\% and 3.0\% over \texttt{LLM4SZZ+}. The results highlight the effectiveness of majority voting in mitigating individual tool weaknesses.

\textbf{Best-in-Dimension Strategy.}
Focusing on patch modification types (Add-only, Del-only, Mixed), the combination of \texttt{SEM-SZZ}, \texttt{LLM4SZZ}, and \texttt{V-SZZ} achieves 50.3\% accuracy at the vulnerability level and 81.7\% F1-score at the version level—surpassing \texttt{VCCFinder} by 5.4\% and 3.9\%. While marginally below \texttt{LLM4SZZ+}, it shows the value of aligning tool selection with orthogonal performance dimensions.

In summary, the experimental results across all three strategies consistently demonstrate the benefits of tool combination.
First, ensemble methods provide substantial improvements over the best standalone and hybrid tools, confirming the overall effectiveness of tool integration.
Second, the top-performing combinations in the Inclusion and Voting strategies consistently integrate both tracing-based and matching-based approaches, validating their complementary strengths.
Third, \texttt{LLM4SZZ+} is included in all leading combinations, underscoring its robustness as a foundation for ensemble configurations.
However, the best vulnerability-level accuracy remains below 60\%, reflecting fundamental limits in existing tool architectures—such as reliance on heuristics, insufficient semantic modeling, and coarse-grained decision logic.

\begin{tcolorbox}[colback=black!10!white,enhanced,frame hidden, boxsep=0pt,left=5pt,right=5pt,top=2pt,bottom=2pt]
\finding
Ensemble strategies significantly improve performance over individual tools, yet the bottleneck highlights common architectural limitations that remain unresolved.
\end{tcolorbox}

\section{Discussion}\label{discussion}

\subsection{Challenges and Implications}

Our study reveals that existing tools for vulnerability-affected version identification suffer from substantial limitations across three dimensions: \ding{182} noisy and coarse-grained patches, \ding{183} semantic mismatch between fix locations and root causes, and \ding{184} shallow presence verification. These issues span diff-based heuristics (e.g., \texttt{V-SZZ}), semantic analysis (e.g., \texttt{MOVERY}), and LLM-based approaches (e.g., \texttt{LLM4SZZ}).

\textbf{Noisy patches.} Most tools assume that any changes in a vulnerability patch are relevant. However, patches frequently contain unrelated edits, such as refactoring or multi-issue fixes, which impair downstream analysis. Our manual inspection indicates that 19\% of patches contain unrelated changes. Despite this, few tools perform effective preprocessing to filter noise. Although \texttt{LLM4SZZ} attempts to isolate relevant edits using large language models, its failure rate remains 35\%, highlighting this task's difficulty.

\textbf{Semantic mismatch.} A more fundamental limitation arises from the disconnect between patch locations and vulnerability root causes. Certain vulnerability types (e.g., \textit{double free}, \textit{use-after-free}) involve complex inter-procedural interactions, with contributing code located in different functions or files. Even within a single function, control-flow-preserving changes (e.g., replacing \texttt{return} with \texttt{goto} for cleanup) can lead to incorrect identification when tools lack flow sensitivity. In our study, \texttt{SEM-SZZ} and \texttt{MOVERY} yielded 39 and 47 false positives, respectively, when identifying vulnerability-related statements across 100 samples.

\textbf{Simplistic presence verification.} Existing tools often rely on tracing heuristics or feature-matching strategies to determine whether a version is vulnerable. Tracing-based methods are prone to misidentifying vulnerability-introducing commits, while feature-matching approaches focus on syntactic similarity without verifying whether the vulnerability condition persists. The absence of root-cause-aware reasoning leads to both high false-positive and false-negative rates.

\subsection{Practical Workarounds via Tool Combination}

Given that no individual tool achieves more than 44.9\% accuracy or 77.8\% F1-score, we explore ensemble strategies as practical alternatives. A majority-voting approach across five tools improves accuracy and F1-score by 10.1\% and 7.0\%, respectively, over the best standalone method. The finding demonstrates that tools offer complementary strengths, and that combination helps offset single tool's limitations. Though not a substitute for deeper technical advances, such ensembles provide a viable interim solution in practice.

\subsection{Future Directions}

To address the above limitations more fundamentally, we highlight three promising research directions:

\textbf{\textit{1) Patch preprocessing.}} Future approaches should move beyond treating patches as atomic units. Combining static analysis with LLMs could enable finer-grained filtering of unrelated edits. Patch modularization—partitioning edits by functional or semantic unit—may further improve the mapping between fixes and individual vulnerabilities.

\textbf{\textit{2) Root cause localization.}} While root cause analysis remains difficult when limited to patches, progress can be made by designing analysis strategies tailored to specific vulnerability types and remediation patterns. Additionally, prompting LLMs with structured expert knowledge—e.g., through chain-of-thought reasoning or multi-role dialogue—may help elicit more accurate vulnerability semantics.

\textbf{\textit{3) Presence verification.}} As code evolves, the boundary between fixed and vulnerable states becomes blurred. We advocate for verification strategies grounded in vulnerability semantics, ideally aligned with identified root causes. Few-shot prompting, multi-agent validation, and root-cause-aware reasoning are promising techniques to improve robustness and reduce both false positives and false negatives.

\subsection{Threats to Validity}

\subsubsection{\textbf{External Validity}}
Our study focuses on C/C++ open-source projects, which may limit the generalizability to other languages and ecosystems. Nonetheless, C/C++ is central to memory-related vulnerabilities, the main focus of existing tools. To ensure coverage, we selected nine actively maintained projects from diverse domains (e.g., OSs, databases, web servers). While our dataset includes 1,128 vulnerabilities, it may underrepresent rare or poorly documented types. We mitigated this by validating coverage against the CWE Top 25, though long-tail cases may still be missing.

\subsubsection{\textbf{Internal Validity}}
Manual labeling of ground truth may introduce error, but it is necessary due to the absence of standard benchmarks. We mitigated this via double annotation by experienced reviewers, with conflicts resolved by a third expert. While tool selection may miss obscure methods, we combined comprehensive keyword searches with multi-round snowballing, covering all major technique categories and emphasizing tools with public implementations. Tools were run with default configurations, which may affect fairness; however, these typically reflect optimal settings as reported. Lastly, manual FP/FN inspection may involve bias; to reduce this, we used standardized criteria and cross-checking, with minimal disagreements resolved through discussion.

\section{Conclusion}

We conducted the first comprehensive empirical study on vulnerability-affected versions identification, evaluating 12 tools across 1,128 real-world C/C++ vulnerabilities. Our results show that no tool achieves over 45.0\% accuracy, largely due to heuristic reliance, limited semantic understanding, and inflexible matching strategies. Even with ensemble methods, accuracy remains below 60.0\%, indicating fundamental limitations in existing designs. Our benchmark and analysis provide practical guidance for tool development and establish a foundation for future work toward more accurate and resilient approaches.

\section*{Acknowledgment}
This research is supported by the National Research Foundation, Singapore, and DSO National Laboratories under the AI Singapore Programme (AISG Award No: AISG2-GC-2023-008); by the National Research Foundation Singapore and the Cyber Security Agency under the National Cybersecurity R\&D Programme (NCRP25-P04-TAICeN); and by the Prime Minister’s Office, Singapore under the Campus for Research Excellence and Technological Enterprise (CREATE) programme.
Any opinions, findings and conclusions, or recommendations expressed in these materials are those of the author(s) and do not reflect the views of National Research Foundation, Singapore, Cyber Security Agency of Singapore, Singapore.
% \newpage

%-------------------------------------------------------------------------------
\newpage
\bibliographystyle{IEEEtran}
\bibliography{ref}

% \vspace{12pt}
% \color{red}
% IEEE conference templates contain guidance text for composing and formatting conference papers. Please ensure that all template text is removed from your conference paper prior to submission to the conference. Failure to remove the template text from your paper may result in your paper not being published.

%%%%%%%%%%%%%%%%%%%%%%%%%%%%%%%%%%%%%%%%%%%%%%%%%%%%%%%%%%%%%%%%%%%%%%%%%%%%%%%%
\end{document}